\documentclass{ar-1col}

\usepackage[comma]{natbib}
\usepackage{url,color,amssymb,subfigure}
\usepackage[dvipsnames]{xcolor}
\usepackage{gensymb}
\usepackage{amsmath}
\usepackage{comment}

\newcommand{\rp}{{\it r}-process\,}
\newcommand{\gw}{GW170817\,}
\newcommand{\grb}{GRB\,170817A\,}
\newcommand{\gfo}{AT\,2017gfo\,}
\newcommand{\ye}{\ensuremath{Y_e}\,}
\newcommand{\xlan}{\ensuremath{X_{\rm lan}}\,}

\newcommand{\Ekjet}{$10^{51}\,\rm{erg}$\,}
\newcommand{\thetajet}{$2\degree-4\degree$\,}
\newcommand{\thetaobs}{$14\degree-19\degree$\,}

\newcommand{\Ekdensityratio}{$(1.5-1.9)\times10^{53}\,\rm{erg\, cm^{-3}}$\,}
\newcommand{\thetaobsthetajetratio}{$5-6$\,}

\newcommand{\tpeak}{$160\,\rm{d}$\,}
\newcommand{\alpharise}{0.8}
\newcommand{\alphadecay}{2.2}
\newcommand{\pelectron}{$2.166 \pm 0.026$\,}
\newcommand{\betaradioX}{0.583 \pm 0.013\,}

\newcommand{\delayGammaRaysGBM}{$1.74\pm0.05$\,s\,} 
\newcommand{\TninetyGBM}{$2.0\pm0.5$\,s\,} 
 
\newcommand{\EgammaisototalGBM}{$(4.8\pm0.9)\times 10^{46}\rm{erg}$\,}
\newcommand{\LpkisototalGBM}{$(1.4\pm0.5)\times 10^{47}\rm{erg\,s^{-1}}$\,}
\newcommand{\EpkfirstGBM}{$185\pm62$\,keV\,} 

\newcommand{\EgammaisofirstGBM}{$(3.6\pm0.9)\times 10^{46}\rm{erg}$\,}
\newcommand{\durationfirstGBM}{$\sim0.5$\,s\,}
\newcommand{\TblackbodyGBM}{$10.3\pm1.5$\,keV\,} 
\newcommand{\durationsecondGBM}{$\sim1.12$\,s\,}
\newcommand{\EgammaisosecondGBM}{$(1.2\pm0.3)\times 10^{46}\rm{erg}$\,}
\newcommand{\offsetarcsec}{$10.315'' \pm 0.007 ''$} 
\newcommand{\offsetkpc}{$2.162\pm 0.001\,\rm{kpc}$} 

\newcommand{\offsetrenormalized}{0.6} 

\newcommand{\HGstellarmass}{$10.65^{+0.03}_{-0.03}$} 
\newcommand{\HGSFR}{$-2.00^{+0.44}_{-0.66}$}  
\newcommand{\HGstellarage}{$11.2^{+0.7}_{-1.4}\,\rm{Gyr}$} 
\newcommand{\HGlastSFRepoch}{$6.8^{+2.2}_{-0.8}\,\rm{Gyr}$} 
\newcommand{\HGmergerepoch}{$\lesssim1\,\rm{Gyr}$} 
\newcommand{\HGXraylum}{$L_X\approx (2-3)\times 10^{39}\,\rm{erg\,s^{-1}}$ (0.5--8 keV)} 
\newcommand{\HGradiolum}{$L_{\rm{6GHz}}\approx 7\times 10^{26}\,\rm{erg\,s^{-1}}$} 

\makeatletter
\let\jnl@style=\rmfamily 
\def\ref@jnl#1{{\jnl@style#1}}%
\newcommand\aj{\ref@jnl{A.~J.}}
\newcommand\araa{\ref@jnl{ARA\&A}}
\newcommand\apj{\ref@jnl{Ap.~J.}}
\newcommand\apjl{\ref@jnl{ApJL}}     
\newcommand\apjs{\ref@jnl{ApJS}}
\newcommand\ao{\ref@jnl{ApOpt}}
\newcommand\apss{\ref@jnl{Ap\&SS}}
\newcommand\aap{\ref@jnl{Astron. Astrophys.}}
\newcommand\aapr{\ref@jnl{A\&A~Rv}}
\newcommand\aaps{\ref@jnl{A\&AS}}
\newcommand\azh{\ref@jnl{AZh}}
\newcommand\baas{\ref@jnl{BAAS}}
\newcommand\icarus{\ref@jnl{Icarus}}
\newcommand\jrasc{\ref@jnl{JRASC}}
\newcommand\memras{\ref@jnl{MmRAS}}
\newcommand\mnras{\ref@jnl{MNRAS}}
\newcommand\pra{\ref@jnl{PhRvA}}
\newcommand\prb{\ref@jnl{PhRvB}}
\newcommand\prc{\ref@jnl{PhRvC}}
\newcommand\prd{\ref@jnl{PhRvD}}
\newcommand\pre{\ref@jnl{PhRvE}}
\newcommand\prl{\ref@jnl{PhRvL}}
\newcommand\pasp{\ref@jnl{PASP}}
\newcommand\pasj{\ref@jnl{PASJ}}
\newcommand\qjras{\ref@jnl{QJRAS}}
\newcommand\skytel{\ref@jnl{S\&T}}
\newcommand\solphys{\ref@jnl{SoPh}}
\newcommand\sovast{\ref@jnl{Soviet~Ast.}}
\newcommand\ssr{\ref@jnl{SSRv}}
\newcommand\zap{\ref@jnl{ZA}}
\newcommand\nat{\ref@jnl{Nature}}
\newcommand\iaucirc{\ref@jnl{IAUC}}
\newcommand\aplett{\ref@jnl{Astrophys.~Lett.}}
\newcommand\apspr{\ref@jnl{Astrophys.~Space~Phys.~Res.}}
\newcommand\bain{\ref@jnl{BAN}}
\newcommand\fcp{\ref@jnl{FCPh}}
\newcommand\gca{\ref@jnl{GeoCoA}}
\newcommand\grl{\ref@jnl{Geophys.~Res.~Lett.}}
\newcommand\jcp{\ref@jnl{JChPh}}
\newcommand\jgr{\ref@jnl{J.~Geophys.~Res.}}
\newcommand\jqsrt{\ref@jnl{JQSRT}}
\newcommand\memsai{\ref@jnl{MmSAI}}
\newcommand\nphysa{\ref@jnl{NuPhA}}
\newcommand\physrep{\ref@jnl{PhR}}
\newcommand\physscr{\ref@jnl{PhyS}}
\newcommand\planss{\ref@jnl{Planet.~Space~Sci.}}
\newcommand\procspie{\ref@jnl{Proc.~SPIE}}

\newcommand\pasa{\ref@jnl{PASA}}

\makeatother

\jname{Xxxx. Xxx. Xxx. Xxx.}
\jvol{AA}
\jyear{YYYY}
\doi{10.1146/((please add article doi))}

\begin{document}

\markboth{Margutti \& Chornock}{GW\,170817}

\title{First Multimessenger Observations of
a Neutron Star Merger}

\author{Raffaella Margutti$^1$ and Ryan Chornock$^{2,1}$
\affil{$^1$Center for Interdisciplinary Exploration and Research in Astrophysics (CIERA) and Department of Physics and Astronomy, Northwestern University, Evanston, IL 60208, USA; email: raffaella.margutti@northwestern.edu}
\affil{$^2$Astrophysical Institute, Department of Physics and Astronomy, 251B Clippinger Lab, Ohio University, Athens, OH 45701, USA; email: chornock@northwestern.edu}}

\begin{abstract}
We describe the first observations of the same celestial object with gravitational waves and light. \\
$\bullet$ \gw was the first detection of a  neutron star merger with gravitational waves. \\
$\bullet$ The detection of a spatially coincident weak burst of $\gamma$-rays (GRB~170817A) 1.7~s after the merger constituted the first electromagnetic detection of a gravitational wave source and established a connection between at least some cosmic short gamma-ray bursts (SGRBs) and binary neutron star mergers. \\
$\bullet$ A fast-evolving optical and near-infrared transient (\gfo) associated with the event can be interpreted as resulting from the ejection of $\sim$0.05~M$_{\odot}$ of material enriched in \rp elements, finally establishing binary neutron star mergers as at least one source of \rp nucleosynthesis. \\
$\bullet$ Radio and X-ray observations revealed a long-rising source that peaked $\sim$\tpeak after the merger. Combined with the apparent superluminal motion of the associated VLBI source, these observations show that the merger produced a relativistic structured jet whose core was oriented $\approx$ 20\degree\, from the line of sight and with properties similar to SGRBs. The jet structure likely results from the jet interaction with the merger ejecta. \\
$\bullet$ The electromagnetic and gravitational wave information can be combined to produce constraints on the expansion rate of the universe and the equation of state of dense nuclear matter. These multimessenger endeavors will be  a major emphasis for future work.
\end{abstract}

\maketitle

\tableofcontents

\section{INTRODUCTION}

Disparate threads of research from astrophysics, general relativity, and nuclear physics were united on 2017 August 17 with the discovery of \gw  by the Advanced Laser Interferometer Gravitational-wave Observatory (LIGO; \citealt{AasiAdvancedLIGO}) and Advanced Virgo \citep{AcerneseAdvancedVirgo} interferometers. A gravitational wave (GW) source with a signal indicative of a compact binary merger \citep{gw170817discovery} was followed by a short gamma-ray burst (SGRB;  \citealt{Goldstein17,Savchenko17}) and an optical counterpart localized to the outskirts of the nearby galaxy NGC~4993 \citep{coulter17}. The intensive campaign across the electromagnetic (EM) spectrum to characterize the source  \citep{Abbott17MMA} was a watershed event in astrophysics, marking the first multimessenger detection of a binary neutron star (BNS) merger.
 Here we review the observations of this event and summarize the key inferences that followed.

\subsection{Thread 1: GW sources and NS mergers}
\label{SubSec:mergers}

Ever since the discovery of PSR 1913+16 \citep{hulse75}, it has been known that compact object binaries exist in our Galaxy, and that at least some of them will decay by emission of gravitational waves (GWs) to merge in less than a Hubble time \citep{taylor82}. Current estimates are that the rate of BNS mergers in the Galaxy is $\mathfrak{R}_{\rm BNS} = 37^{+24}_{-11}$~Myr$^{-1}$ (90\% confidence; \citealt{pol20}).
The existence of GW emission from compact object binaries was spectacularly confirmed by the detection of the binary black hole (BH) system GW150914 \citep{abbott150914discovery}.

What happens in a merger when one of the compact objects is a neutron star (NS)? \citet{ls74} first considered NS-BH mergers and found that $\sim$5\% of the NS material might be ejected.  This idea was soon extended to BNS mergers \citep{sym82,Eichler89}. Simulations have grown in numerical sophistication over the last few decades and identified
several possible mass ejection mechanisms in BNS mergers \citep{rosswog99,oechslin07,Bauswein13,Hotokezaka13}. In the last few orbits before the merger occurs, the NSs are tidally squeezed and eject tidal tails of decompressed NS material \citep[e.g.,][]{Sekiguchi16}. Further dynamic ejection of matter occurs at the collision interface between the two NSs \citep[e.g.,][]{wanajo14}. Up to 0.1~M$_{\odot}$ of material can then form a disk around the remnant compact object \citep{Radice18BNS}. Winds and outflows driven from either a hypermassive NS (HMNS) remnant or the disk surface can be powered by neutrinos, magnetic fields, or viscous effects, resulting in enhanced mass loss \citep{dessart09,metzger14bluekn,perego14,siegel14}.
The dynamics of the merger process and mass ejection have been reviewed recently by \citet{shibata19arnps} and \citet{Radice20annurev}.

\subsection{Thread 2: The Mystery of the \rp}

Nuclei of elements heavier than the iron peak are primarily formed through neutron capture reactions. \citet{bbfh} and \citet{cameron57} identified two separate processes that were necessary to produce the measured isotopic ratios of the heavy elements in the solar system \citep{suessurey}. The key distinction is whether the timescale between successive neutron captures is substantially shorter or longer than the beta decay timescales of the unstable nuclei that form. While the slow {\it s}-process has been conclusively shown to occur in evolved stars, the site(s) of the rapid \rp\ have been debated for more than sixty years \citep[e.g.,][]{sneden08,cowan19}. 
In parallel with these debates, significant constraints were provided by studies of the abundances of neutron capture elements in extremely metal-poor stars (e.g., \citealt{frebel18}). They have shown that the abundance ratios for the heaviest \rp\ elements (Ba and higher) were very close to those seen in the Sun, while the lighter \rp\ elements exhibited significantly higher scatter from star to star.
The initial suggestion was that supernovae and the vicinities of the newly formed NSs in their interiors provided the hot, neutron-rich environments favorable for the \rp. However, detailed numerical simulations of the resulting nucleosynthesis have generally failed to reliably produce the heaviest elements in sufficient abundance \citep[e.g.,][]{qian96}. 

The neutron-rich ejecta of NS-BH mergers were suggested to be interesting potential sites for \rp\ nucleosynthesis \citep{ls74,ls76}, as were BNS systems \citep{sym82,Eichler89}. 
Detailed nucleosynthesis calculations of BNS mergers supported the idea that they might result in significant \rp production \citep{freiburghaus99,rosswog99}. An important parameter is the electron fraction, \ye, which is the ratio of the number of protons to nucleons in the ejecta material. Low \ye material is neutron rich and can experience the \rp. However, doubts remained about whether the event rates or ejecta masses were sufficiently high to contribute significantly to Galactic nucleosynthesis \citep[e.g.,][]{qw07}.
Nonetheless, circumstantial evidence from such disparate lines of evidence as low $^{244}$Pu abundances in the ocean floor \citep{hotokezaka15,wallner15} and a large \rp\ enhancement in the chemical abundances of the dwarf galaxy Reticulum II \citep{ji16} accumulated and pointed to rare events with large yields being responsible for the majority of \rp\ production, particularly for the heaviest neutron-capture elements, which was hard to accommodate in supernova models. 
Resolution of the \rp\ mystery would also give insight into the nuclear physics of neutron-rich isotopes, only some of which are currently experimentally accessible \citep{mumpower16,horowitz19}.

\subsection{Thread 3: SGRBs} \label{SubSec:thread3SGRBs}
Short-duration $\gamma$-ray bursts (SGRBs) are cosmic flashes of $\gamma$-rays \citep{klebesadel73} with durations of less than 2~s \citep{kouveliotou93}.
Circumstantial observational evidence of the association of SGRBs  with mergers of compact objects (either NS-NS or NS-BH) has been reviewed by \cite{Berger14} and includes very deep optical observations that rule out the presence of supernova explosions; observed optical/near-infrared (NIR) excesses of emission with respect to the afterglow decay, detected in some nearby SGRBs with properties consistent with \rp powered kilonovae; the remote locations of SGRBs in their host galaxies, which are of early-type morphology in $\sim 1/3$ of events, indicative of older stellar populations; additionally, the location of SGRBs within their host galaxies is weakly correlated with the underlying host-galaxy light distribution, indicating that SGRBs are not good tracers of star formation or stellar mass. These indirect pieces of evidence collectively favor NS-NS/NS-BH mergers as progenitors of SGRBs, as was theoretically postulated by \citet{Eichler89} and \citet{Narayan92}. Yet, before GW\,170817, no  \emph{direct} observational evidence supported this conclusion.

\subsection{A (Brief) Overview of EM Counterpart Predictions}
\label{subsec:counterparts}

Prior to \gw, EM counterparts to NS mergers detected through GWs had been proposed across the spectrum from $\gamma$-rays to radio, with a range of timescales from seconds to years (see the review by \citealt{fernandez16review}). The odds that the narrow $\gamma$-ray emitting relativistic jet would be aimed directly at the observer were regarded as low for any individual event, despite the importance of making such a connection \citep{metzger12}. This motivated a focus on more isotropic signatures for counterpart searches.

\citet{li98} realized that the combination of NS merger ejecta unbound to the final compact object with a source of energy from radioactive decays should result in an observable optical transient, although they predicted high luminosities ($L_{\rm peak}\approx 10^{44}$~erg~s$^{-1}$) and short durations ($\sim$1~d). These events were sometimes referred to as ``mini-supernovae". \citet{kulkarni05} considered counterparts powered by decay of free neutrons or nickel and introduced the term ``macronova".  The first models to incorporate more realistic treatments of radioactive heating and thermalization by \rp decay products assumed simplified opacities inspired by Thomson scattering and iron-peak elements  \citep{metzger10,goriely11,roberts11}. As the significantly lower predicted luminosities of $L_{\rm peak}\approx 10^{41}$--10$^{42}$~erg~s$^{-1}$ were approximately 1000$\times$ those of classical novae, \citet{metzger10} coined the term ``kilonova"  for these transients, which we adopt here. 

A major theoretical breakthrough was provided by considerations of the opacities of lanthanide elements, which are copiously produced by the strong \rp \citep{kasen13}. Atoms and ions whose valence electrons partially fill the $f$-shell have a substantially larger number of low-lying energy levels and hence bound-bound transitions available to them than iron-peak elements.  This dramatically increases the opacity of lanthanide-rich material at optical wavelengths, which delays and lowers the peak luminosity of a kilonova and pushes flux to emerge in the NIR, which we will refer to as a ``red" kilonova \citep{Barnes13,kasen13,tanaka13}. Later, it was realized that lanthanide-poor material ejected with high \ye, potentially from polar dynamical ejecta or winds, could produce a ``blue" kilonova that dominates the optical emission \citep{metzger14bluekn}.
The histories of kilonova predictions and evolving input physics have been comprehensively reviewed by \citet{metzger19kn}.
\citet{Nakar11} also predicted that the kilonova ejecta would emit synchrotron radiation as they decelerate in the ambient medium to produce a detectable radio counterpart to BNS mergers on timescales of years.

A solid prediction from the numerical simulations described above is that the merger site is surrounded by 
some  baryon contaminated region with large mass ($\gtrsim0.01\,\rm{M_{\odot}}$).
A SGRB-like jet launched by the merger would thus have to pierce through these outflows of material before breaking out. Before \gw it was realized that the jet propagation 
within the BNS ejecta is a critical step that shapes the jet's final angular structure and collimation, and that determines its ultimate fate (successful vs. choked; e.g.,
\citealt{Aloy05,Bromberg11,Nagakura14,Duffell15,Lazzati2017before170817,Murguia-Berthier17,Nakar17,Gottlieb18cocoonUV}).  While advancing through the ejecta the jet dissipates energy into a hot cocoon (i.e., a wide-angle outflow constituted of shocked jet and ejecta material),   which expands relativistically after breaking out of the ejecta.
Numerical simulations suggest that for standard parameters of successful jets, the time spent by the jet within  the BNS ejecta is comparable to the duration of the subsequent SGRB $\gamma$-ray emission, implying that the cocoon energy and the GRB energy are expected to be similar. Just like the jet, the cocoon has clear electromagnetic signatures associated with it, including cocoon breakout $\gamma$-ray emission, ultraviolet (UV) cooling emission, radioactive heating, and a broadband afterglow. None of these cocoon observational signatures were confidently detected in SGRBs before \gw, and some still remain elusive (\S\ref{Sec:GammaRays}--\ref{Sec:Afterglow}). 

\subsection{The Events of 2017 August 17}

A GW signal was detected as a compact binary coalescence by Advanced LIGO  and Advanced Virgo  on 2017 August 17, with the end of the inspiral signal at 12:41:04.4 UTC (after $\sim100$~s of GW emission detectable in the LIGO band; \citealt{gw170817discovery}). This event was followed \delayGammaRaysGBM later by a burst of $\gamma$-rays detected by {\it Fermi}-GBM \citep{Goldstein17} and INTEGRAL SPI-ACS \citep{Savchenko17}.
The subsequent detection of an optical counterpart to \gw and \grb was first made by the One-Meter Two-Hemispheres collaboration using the Swope telescope at 10.9~hr after the merger \citep{coulter17}, and within the next hour five other teams independently detected the same source, now known as \gfo \citep{Arcavi17,lipunov17,soares17,Tanvir17,valenti17}. No X-ray or radio counterpart was detected down to deep limits during the first few days of observations \citep{Alexander17,Evans17,Hallinan17,Savchenko17,Margutti17GW,Sugita18}.
A rising X-ray and radio source eventually crossed the threshold of detection of sensitive X-ray (CXO) and radio (VLA) observatories on day 8.9 \citep{Troja17} and day 16.4 \citep{Hallinan17} of monitoring, respectively (with a tentative radio detection at $\delta t=10.4$~d).
\citet{Abbott17MMA} give a detailed account of the timeline of EM observations that followed the initial detection of the GW source. Multimessenger observations of \gw{} and their implications have also been reviewed by \cite{Nakar19review,Burns19review}.
\begin{textbox}[t]\section{Nomenclature}
The electromagnetic counterpart to \gw\ has been referred to by several equivalent names in the literature: SSS17a \citep{coulter17}, DLT17ck \citep{valenti17}, MASTER OTJ$130948.10-232253.3$ \citep{lipunov17}, and EM170817 \citep[e.g.,][]{Evans17,Kasliwal17}.
We adopt the official International Astronomical Union transient name, \gfo, in this review for the thermal UV/optical/IR source. When speaking specifically of the burst of  $\gamma$-rays, which was automatically labeled \grb, we follow the convention from that community. We refer to the GW event as \gw, and to the associated non-thermal emission as synchrotron emission from the afterglow of \gw.
\end{textbox}

Several different techniques give redshift-independent distance estimates of $\sim$40~Mpc for the host galaxy NGC\,4993 (e.g., \citealt{hjorth17}). 
Here we adopt 
$D_{L}=40.7\pm 1.4 \pm 1.9\,\rm{Mpc}$ (random and systematic uncertainties, respectively)   obtained from surface brightness fluctuations in NGC 4993 \citep{cantiello18}. Time is referenced to the GW coalescence time, 2017-08-17 12:41:04.4 UTC (or MJD=57982.528523), \cite{gw170817discovery}. All photometry and spectroscopy in the figures have been corrected for $E(B-V)=0.105$~mag of Galactic reddening \citep{Schlafly11}.  We adopt the MUSE/Very Large Telescope measurement of the heliocentric redshift of NGC 4993 $z_{helio}=0.0098$ \citep{hjorth17,Levan17}, which gives a geocentric redshift $z=0.0099$ at the time of \gw. All quoted magnitudes are on the AB system. Uncertainties (upper limits) are provided at the $1\,\sigma$ ($3\,\sigma$) Gaussian-equivalent confidence level unless explicitly mentioned otherwise.

\section{GW170817: GRAVITATIONAL WAVE EMISSION} \label{Sec:GWs}
The GW signal of \gw{} was detected at high significance in both the Advanced LIGO-Hanford and LIGO-Livingston detectors (combined signal-to-noise ratio of 32.4, with a false alarm rate of less than one in $8\times10^4$~yr; \citealt{gw170817discovery}), despite a detector glitch in the LIGO-Livingston detector that appeared 1.1~s prior to coalescence. This event only resulted in a signal-to-noise ratio of $\sim$2 in the Virgo detector, surprisingly low given the proximity of the event, and which implied a localization near the detector nulls. The  luminosity  distance estimate was $D_{L}=40^{+8}_{-14}\,\rm{Mpc}$ and 
the skymap had a 90\% GW localization region of 28~$\deg^2$ \citep{gw170817discovery}, which was later reduced to 16~$\deg^2$ after reanalysis \citep{abbott19properties}. Both regions include the sky location of \gfo, and the GW localization volume includes its host galaxy, NGC\,4993.

\begin{figure}[h]
\hskip -1. cm
\includegraphics[width=5.5in]{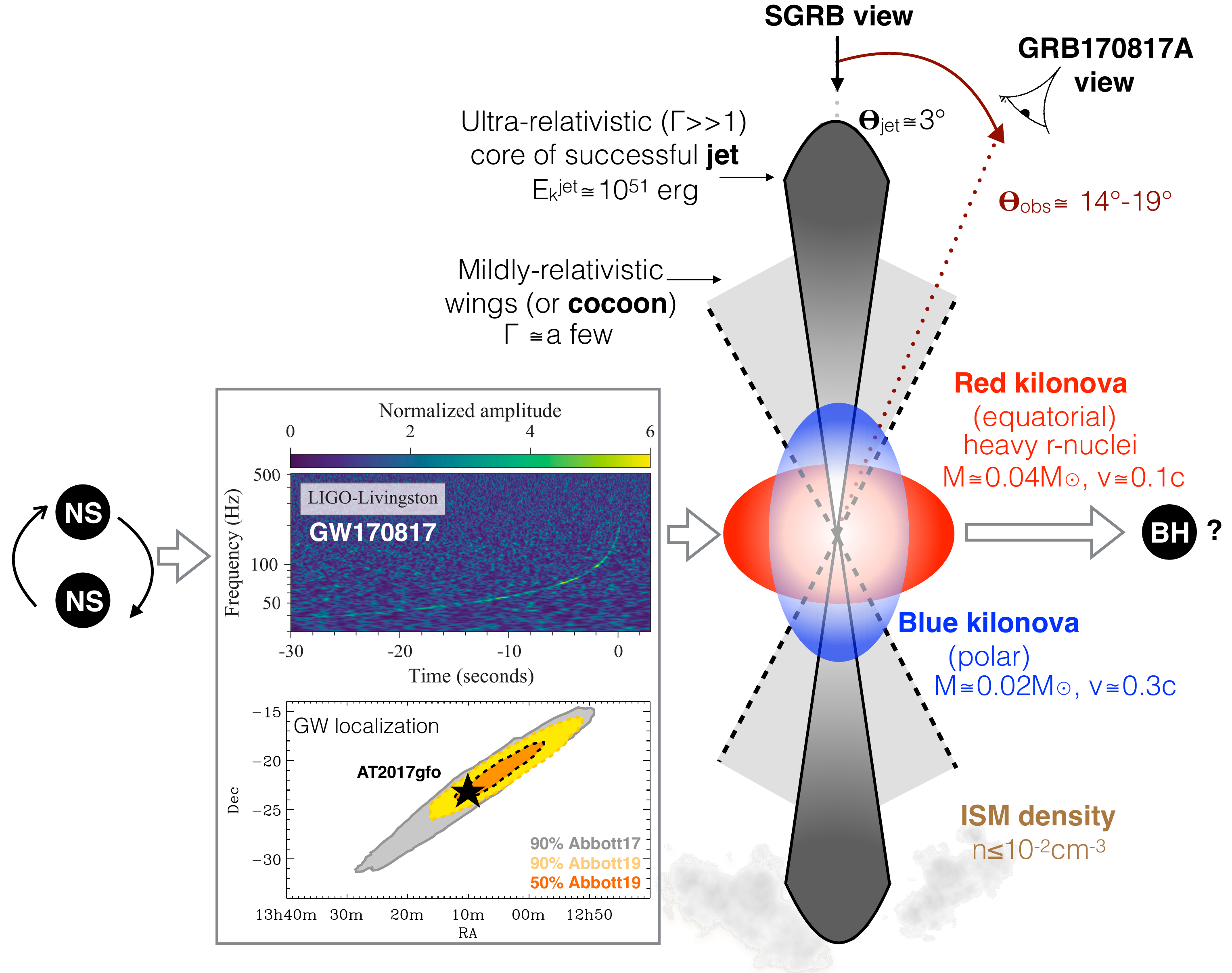}
\caption{Composite panel and cartoon showing the EM+GW emission components and geometry of the compact-object merger \gw{}. The GW emission was detectable until the merger time ($t=0$ in the time-frequency representation of the LIGO-Livingston data, upper panel, adapted  from \citealt{gw170817discovery}), and enabled an initial localization within 28 deg$^2$ \citep{gw170817discovery}, later refined to 16 deg$^2$ (\S\ref{Sec:GWs}, \citealt{abbott19properties}). The merger produced a burst of $\gamma$-rays (GRB\,170817A, \S\ref{Sec:GammaRays}), and a multi-wavelength afterglow powered by a collimated relativistic jet with a wider-angle component of mildly relativistic material viewed off-axis (\S\ref{Sec:Afterglow}). A multi-color kilonova dominated the UV-optical-NIR spectrum for the first weeks (\S\ref{Sec:Thermal}). While the ultimate fate of the merger remnant cannot be probed directly, indirect evidence favors a BH (\S\ref{SubSec:RemnantNature}). }
\label{Fig:cartoon}
\end{figure}

\subsection{Source parameters and basic inferences} \label{SubSec:GWparameters}
The physical properties of GW sources are inferred by matching the observed data with waveforms generated following the prescriptions of general relativity, which makes detailed predictions for the inspiral and coalescence signal of merging NSs and BHs.
The observed waveform depends on a combination of intrinsic and extrinsic parameters. Intrinsic parameters include the component masses $m_1$ and $m_2$ (and their notable combinations, i.e., mass ratio $q\equiv m_2/m_1\le 1$ and chirp mass $\mathcal{M}$),
\begin{marginnote}[]
\entry{Chirp Mass}{$\mathcal{M}\equiv \frac{(m_1m_2)^{3/5}}{(m_1+m_2)^{1/5}}$. It is the best measured parameter from the modeling of GW sources with a long inspiral.} 
\end{marginnote}
spin angular momenta of the two bodies that contribute six parameters $\mathcal{S}_i$ (typically expressed in terms of their dimensionless forms $\chi_i$), and parameters governing the tidal deformability of each binary component.
\begin{marginnote}[]
\entry{Dimensionless Spin}{ $\chi_i\equiv \frac{c\mathcal{S}_i}{(G\,m_i^2)}$,  where $\mathcal{S}_i$ is the $i$ spatial component of the spin angular momentum. }
\end{marginnote}
Extrinsic parameters are those related to the localization of the GW event in the sky, the luminosity distance, and the orientation of the binary angular momentum with respect to the observer ($\theta_{JN}$). The constraints on the deformability of matter derived from this system are discussed in \S\ref{SubSec:deformability}. Here we focus on the component masses and spins.

A first estimate of these parameters was presented by \cite{gw170817discovery}. \cite{abbott19properties} present more precise parameter constraints that primarily result from  the reduced calibration uncertainties of Virgo data, a broader bandwidth of GW data included in the analysis (extending to 23~Hz compared to the 30~Hz of the original analysis of \citealt{gw170817discovery}, which gives access to additional $\sim$1500 cycles out of a total of $\sim$3000), a wider range of improved waveform models, and knowledge of the source location within NGC\,4993 from EM observations.   By making minimal assumptions about the nature of the merging compact objects, and, specifically, by allowing for a large range of deformabilities that include the possibility of BH components, \cite{abbott19properties} derive a primary mass $m_1 \in (1.36,1.89)\,\rm{M_{\odot}}$ (one-sided 90\% lower and upper limit), a secondary mass $m_2\in (1.00,1.36)\,\rm{M_{\odot}}$ and a total mass of $2.77^{+0.22}_{-0.05}\,\rm{M_{\odot}}$ (median, 5\% lower-limit and 95\% upper limit) enforcing a prior $\chi\le0.89$. Individual components spins are less well constrained to $\chi_1\in(0.00,0.50)$ and $\chi_2\in(0.00,0.61)$. The mass values are consistent with being drawn from the Galactic NS population (which is described by a mean value of $m_{\rm{MW}}=1.32\,\rm{M_{\odot}}$ with standard deviation $\sigma_{\rm{MW}}=0.11\,\rm{M_{\odot}}$, \citealt{Kiziltan13}), yet this comparison does \emph{not} imply that \gw{} contained NSs (\S\ref{SubSec:deformability}). 
The detection of EM emission (\S\ref{Sec:Thermal}-\S\ref{Sec:Afterglow}) however does imply that at least one of the merging compact objects is a NS.
\begin{marginnote}[]
\entry{Binary Inclination Angle $\theta_{JN}$}{Angle between the binary angular momentum and the observer's line of sight.}
\end{marginnote}

\begin{marginnote}[]
\entry{Observer Angle $\theta_{obs}$}{If the jet is launched along the direction of the angular momentum, $\theta_{obs}$ is the minimum between $\theta_{JN}$ and ($180\degree-\theta_{JN}$). }
\end{marginnote}

The fastest spinning Galactic BNSs that will merge within a Hubble time have an extrapolated $\chi\le0.04$ at the time of merger.
A low-spin prior $\chi\le0.05$ consistent with this known Galactic population leads to $m_1 \in (1.36,1.60)\,\rm{M_{\odot}}$, $m_2 \in (1.16,1.36)\,\rm{M_{\odot}}$ and total mass $2.73^{+0.04}_{-0.01}\,\rm{M_{\odot}}$. As expected, in both spin scenarios the chirp mass is derived with much higher precision $\mathcal{M}=1.186^{+0.001}_{-0.001}\,\rm{M_{\odot}}$, with the main source of uncertainty contributing to the quoted error bars being the unknown source velocity in NGC\,4993 (which is quantified by the observed line of sight velocity dispersion of NGC\,4993). Finally, the GW analysis points at a binary system that is inclined with respect to the observer's line of sight of  $\theta_{JN}=151^{\circ +15}_{-11}$  ($\theta_{JN}=153^{\circ +15}_{-11}$) for the low-spin (high-spin) prior after using the $D_{L}$ measurement from EM observations of the host galaxy (see  \S\ref{SubSec:Cosmology}).  
All parameter ranges and uncertainties are quoted following the 90\% convention above.
Adding more informative priors that are astrophysically motivated, and assuming that both objects are NSs has minimal impact on the individual mass estimates, but enables tighter constraints on the deformability of matter of \S\ref{SubSec:deformability} \citep{abbott18eos}. 

\section{GRB\,170817A: $\gamma$-RAY EMISSION} 
\label{Sec:GammaRays}
\begin{figure}[h]
\hskip -1. cm
\includegraphics[width=4.3in]{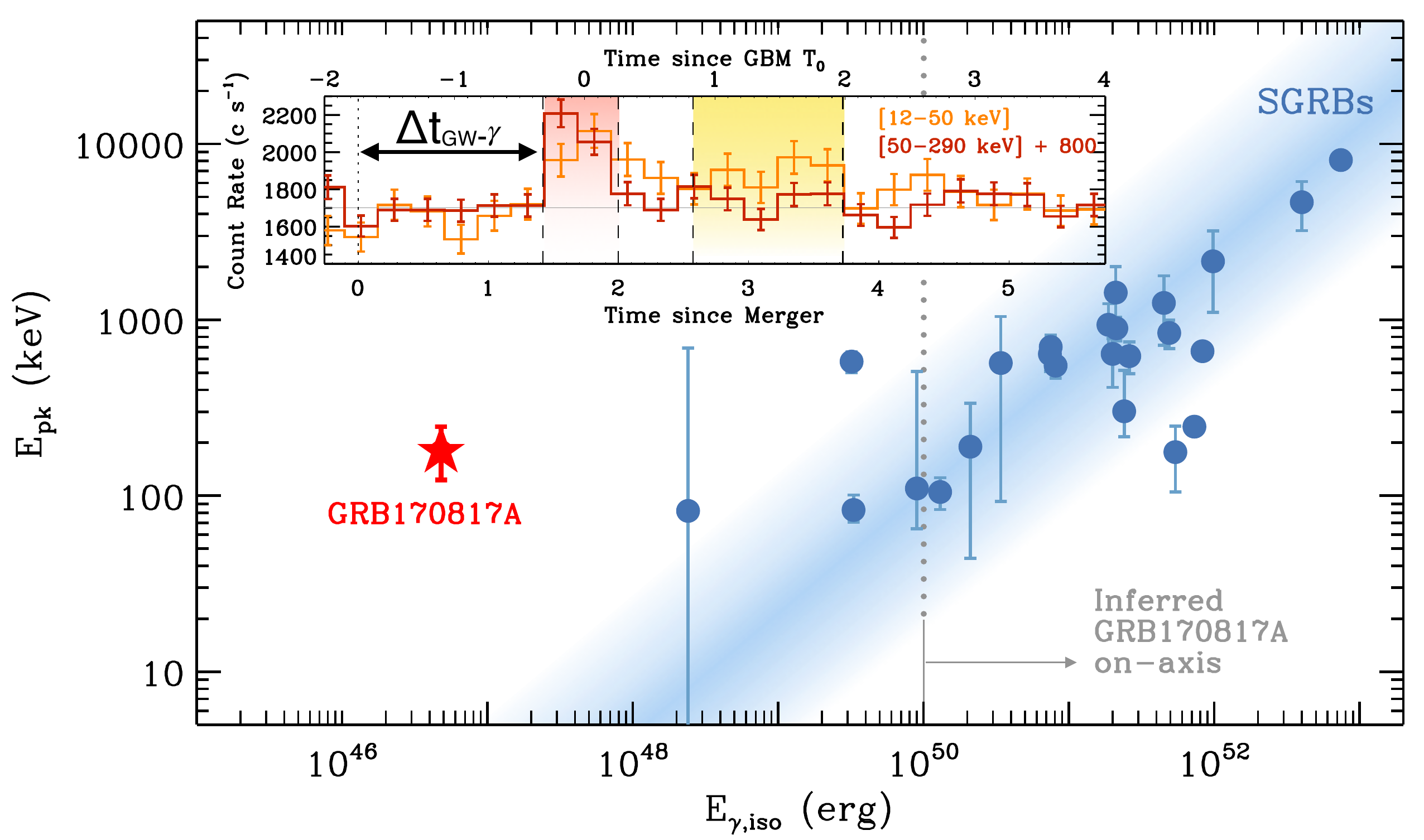}
\caption{With significantly lower  $E_{\gamma,iso}$ but comparable $E_{pk}$, the observed properties of GRB\,170817A (red star) clearly deviate from SGRBs (blue circles, from \citealt{Salafia19}). \emph{Inset:} The {\it Fermi}-GBM light-curve of GRB\,170817A shows a peculiar morphology, with a short hard main pulse of \durationfirstGBM (red shaded area) followed by a softer tail of emission with duration \durationsecondGBM (yellow shaded area). The onset of the $\gamma$-ray emission is delayed compared to the merger time of $\Delta t_{GW-\gamma}$. }
\label{Fig:EpkEiso}
\end{figure}
The burst of $\gamma$-rays detected by {\it Fermi}-GBM \citep{Goldstein17}, and INTEGRAL SPI-ACS \citep{Savchenko17} at $\Delta t_{GW-\gamma}=$\delayGammaRaysGBM after the binary coalescence and spatially coincident with the GW localization of the BNS merger \gw{}, represents the first EM signature physically associated with a GW source (probability of chance coincidence of $5\times 10^{-8}$, or 5.8$\sigma$ Gaussian equivalent, \citealt{abbott17gammaray}) and marks the dawn of multimessenger astrophysics with GWs. 
The key observational properties of the $\gamma$-ray counterpart to \gw are as follows (\citealt{Goldstein17}; see also \citealt{Pozanenko18,Fraija19}). 
The {\it Fermi}-GBM light-curve of GRB\,170817A showed a peculiar morphology consisting of a spike of emission of \durationfirstGBM (also detected by INTEGRAL) followed by a lower-significance tail of softer emission, with total duration of $T_{90}=$\TninetyGBM (\textbf{Figure \ref{Fig:EpkEiso}}). 
The spectrum of the short spike is well fit by a power-law with exponential cutoff (i.e., a Comptonized model) with  peak energy of the  $\nu F_{\nu}$ spectrum $E_{pk}=$\EpkfirstGBM and isotropic equivalent energy release $E_{\gamma,iso}=$\EgammaisofirstGBM (10--1000\,keV).
The spectrum of the softer tail can be fit with a blackbody model with temperature $T=$\TblackbodyGBM and $E_{\gamma,iso}=$\EgammaisosecondGBM, even if the limited photon statistics prevent any conclusive statement about the nature of the intrinsic spectrum. GRB\,170817A showed no evidence for a $\gamma$-ray precursor or extended emission (EE).
\begin{marginnote}[]
\entry{Extended Emission (EE)}{Period of up to $\sim100\,$s of enhanced $\gamma$-ray activity after the short $\gamma$-ray spike that can be energetically dominant 
(as in SGRB\,080503).}
\end{marginnote}

The fact that GRB\,170817A is significantly less energetic than cosmological SGRBs (Fig. \ref{Fig:EpkEiso}) is not surprising, as the most likely scenario of GW-detected BNS mergers is that of an off-axis configuration (typical observer angle $\theta_{obs}\sim30\degree$; 
\citealt{Schutz11}), for which the observed emission is significantly depressed and effectively undetectable \citep{Goldstein17,abbott17gammaray} at  the typical distances and jet-collimation angles of SGRBs  ($z\approx0.5$, $\theta_{jet}\sim 4\degree-15\degree$, \citealt{Berger14,Fong15}).
The true surprise is that the first GW-detected BNS merger 
was \emph{also}  accompanied by the independent detection of $\gamma$-rays.

\begin{figure}[h]
\hskip -1. cm
\includegraphics[width=4in]{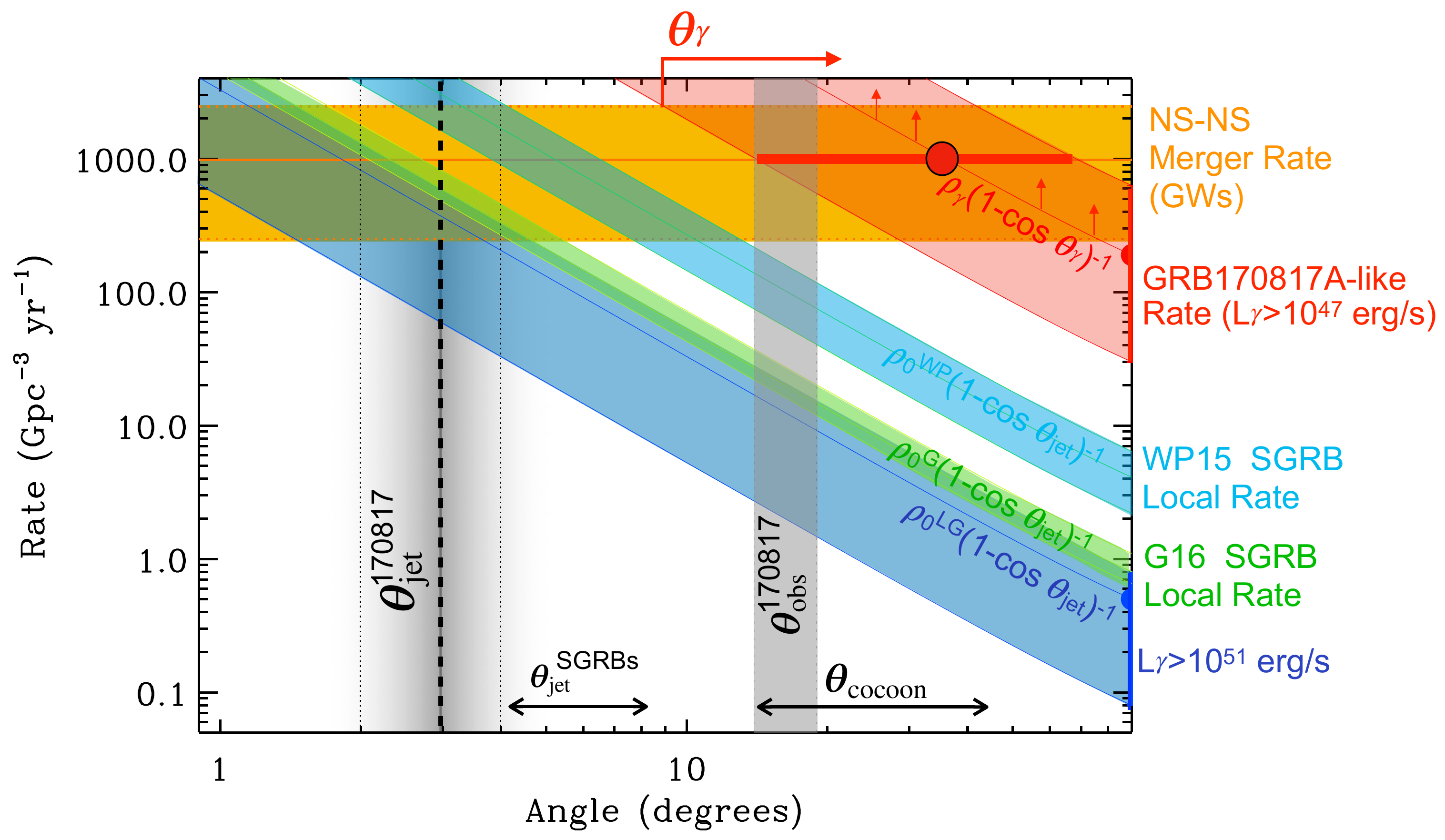}
\caption{Inferred beaming-corrected local rate of \grb-like bursts (red area, \citealt{Zhang18Rate}) and cosmological SGRBs (from \citealt{Wanderman15} and \citealt{Ghirlanda16}, in light-blue and green, respectively) as a function of beaming angle. The local rate of SGRBs with luminosity similar to the inferred on-axis $\gamma$-ray luminosity of GW\,170817 is shown in blue \citep{Ghirlanda19}.  A comparison with 
the
local rate of NS-NS mergers from GW observations (orange area, \citealt{Abbott20latestBNSrate}), reveals that GRB\,170817A likely produced $\gamma$-rays detectable from a relatively large $\theta_{\gamma}>10\degree$, significantly larger than $\theta_{jet}\approx $\thetajet (\S\ref{Sec:Afterglow}) and consistent with the inferred observer location $\theta_{obs}\approx$\thetaobs (\S\ref{Sec:Afterglow}). The range of SGRBs $\theta_{jet}$  jet break 
is indicated with an arrow (\S\ref{Sec:Afterglow}). The range of angular extent $\theta_{cocoon}$ of cocoon outflows/jet wings is also indicated (
e.g., \citealt{Lazzati2017post170817,Gottlieb18gammarays,Kathirgamaraju2019EMcounterparts}). 
}
\label{Fig:Rates}
\end{figure}
\subsection{Relationship to the $\gamma$-ray emission from short GRBs} \label{SubSec:gammaraysSGRBs}

With $E_{\gamma,iso}=$\EgammaisototalGBM and a peak luminosity $L_{pk,iso}=$\LpkisototalGBM,  \grb is orders of magnitude less energetic and luminous than SGRBs, yet with similar duration of $\sim2\,$s. This opens two possibilities: (i) \grb is intrinsically sub-energetic and represents a new class of $\gamma$-ray transients associated with BNS mergers; (ii) \grb is a normal SGRB viewed off-axis. The afterglow observations of \S\ref{Sec:Afterglow} require an off-axis collimated outflow with large energy similar to that of cosmological SGRBs, ruling out the first scenario.
\begin{marginnote}[]
\entry{Top-hat Jet}{Jet with all of the energy uniformly distributed and  confined within $\theta\le \theta_{jet}$.}
\end{marginnote}
In the context of an off-axis view of a relativistic top-hat jet, 
the off-axis correction factor needed to bring the observed $E_{\gamma,iso}$ of \grb in line with that of SGRBs would lead to an extremely large intrinsic peak energy $\gtrsim 3$ MeV \emph{and} to a viewing angle fortuitously just outside the jet ($\theta_{obs}-\theta_{jet}\approx 1\degree$), which is inconsistent with the late afterglow turn on of \S\ref{Sec:Afterglow} \citep{Kasliwal17,Murguia-Berthier17GW170817,abbott17gammaray,Granot17}.  
The conclusion is that the observed $\gamma$-rays require the presence of a \emph{structured} outflow with energy $E > 0$ and Lorentz factor $\Gamma  >1$ outside the jet core at $\theta>\theta_{jet}$. The off-axis view ultimately enabled us to appreciate the presence of this structure, which has a primary role in shaping the peculiar light-curve morphology of \grb, and allowed energetically subdominant components (to which the on-axis SGRB phenomenology is largely insensitive) to emerge. 

We end with a consideration on the energetics of GRB\,170817A in the context of the $\gamma$-ray luminosity function of SGRBs 
\citep{Wanderman15,Sun15,Ghirlanda16}. 
With $L_{pk,iso}=$\LpkisototalGBM, GRB\,170817A extends the luminosity function of SGRBs by more than two orders of magnitude, and implies a current local rate of  $\rho_{\gamma}(L_{pk,iso}>10^{47} \,\rm{erg\,s^{-1}}) \ge 190^{+440}_{-160}\,f_{b\gamma}^{-1}\rm{Gpc^{-3}yr^{-1}}$ \citep{Zhang18Rate}, where $f_{b\gamma}\equiv  (1-cos(\theta_\gamma))$ is the beaming factor and $\theta_{\gamma}$ is the opening angle of the $\gamma$-ray emission.  For typical efficiencies of kinetic energy to $\gamma$-ray emission conversion $\epsilon\approx 10\%-20\%$, the GW\,170817 afterglow energetics of \S\ref{Sec:Afterglow} point at an \emph{on-axis}  $L_{pk,iso}^{on-axis}>10^{51}\,\rm{erg\,s^{-1}}$ \citep{NakarPiran18,Ghirlanda19,Salafia19}, for which the current local rate derived from the SGRB luminosity function is $\rho_{j}(L_{pk,iso}>10^{51} \,\rm{erg\,s^{-1}}) \approx 0.5\,f_{bj}^{-1}\rm{Gpc^{-3}yr^{-1}}$.   A comparison of  $\rho_{\gamma}$ and $\rho_{j}$ assuming that the SGRB luminosity function maps the properties of SGRB jets seen on-axis (at least at the high luminosities $L_{pk,iso}>10^{51} \,\rm{erg\,s^{-1}}$ of interest; e.g., \citealt{Beniamini19}) and that GW\,170817 seen on axis belongs to this distribution leads to $f_{b \gamma} \gtrsim 380^{+7495}_{-343} f_{bj}$ or $\theta_{\gamma} \gtrsim 60^{\circ +30}_{-40}$ for $\theta_{jet}$$\sim$$ 3 \degree$ (\S\ref{Sec:Afterglow}).  Similarly, by requiring the local rate of events with wide-angle $\gamma$-ray emission $\rho_{\gamma}$ to be at most the BNS merger rate inferred from GWs ($\mathfrak{R}=980^{+1490}_{-730}\,\rm{Gpc^{-3}yr^{-1}}$, \citealt{Abbott20latestBNSrate}),  $\theta_{\gamma} \gtrsim 36^{\circ +54}_{-25}$ is inferred, where $\theta_{\gamma}$ is the collimation angle of the $\gamma$-ray emission.  As shown by \textbf{Figure \ref{Fig:Rates}}, current GW and EM observations are thus consistent with the notion that most (if not all) BNS mergers produce 
jets with typical $\theta_{jet}$ of few to ten degrees \citep{Beniamini19} and that all SGRBs might be accompanied by wider angle $\gamma$-ray emission similar to GRB\,170817A. 
Given the low luminosity of this component, it is no surprise that it eluded detection in 
SGRBs. GRB\,170817A-like events would be detectable only out to $\sim$$55$$-$$80$ Mpc with
current 
$\gamma$-ray spacecraft \citep{abbott17gammaray,Goldstein17,Zhang18Rate}.

\subsection{Origin of $\gamma$-rays in this event}
\label{SubSec:GammaRays170817}

The observed delay between the BNS merger and \grb ($\Delta t_{GW-\gamma}=$\delayGammaRaysGBM) is a novel multi-messenger observational parameter, which we assume 
is astrophysical in origin (i.e., GWs and light propagate at the same speed; see also \S\ref{SubSec:GRtests}).
$\Delta t_{GW-\gamma}$ has three major components \citep{Zhang19}: the jet injection time $\Delta t _{inj}$, the jet/cocoon breakout time  $\Delta t_{bo}$, and the time it takes for the emitting material to reach the transparency radius $\Delta t_{tr}$, where the $\gamma$-rays can freely escape. These quantities are not well known from theory and are not well constrained by observations, yet they represent fundamental aspects of the BNS merger physics.   Numerical simulations of SGRB jets breaking through neutrino-driven and magnetically-driven  winds  (\citealt{Murguia-Berthier17}) and dynamical ejecta (\citealt{Nagakura14,Gottlieb18cocoonUV}) 
suggest $\Delta t_{bo}\approx$ few $100$ ms, with high sensitivity on the properties of the ejecta cloud. Observations of $\gamma$-rays from SGRBs suggest similar values $\approx 200-500$ ms \citep{Moharana17}. $\Delta t _{inj}$ ($\approx$10 ms to seconds) depends on the type of jet launching mechanism and likely reflects the timescale needed to form an accretion disk or strong magnetic fields, or the time-scale for the HMNS (if one was formed) to collapse to a BH (e.g., \citealt{Granot17}). Since the properties of the expanding merger ejecta cloud (i.e., density, radius, angular distribution) as seen by the jet strongly depend on how delayed the jet injection is, the structure and  energy partitioning $E(\Gamma \beta)$ within the outflow that result from the jet interaction with the merger's ejecta is also very sensitive to $\Delta t _{inj}$ (e.g., \citealt{Murguia-Berthier20}). 

Compactness arguments applied to the main non-thermal pulse of GRB\,170817A set a limit on the wide-angle outflow Lorentz factor $\Gamma\gtrsim 2.5$ for the source to be optically thin to $e^+e^-$ pair production, which, together with the observed pulse duration, implies a radius of emission $R_{\gamma}=2c\Gamma^2 \delta t_{obs} \gtrsim 10^{11}\,\rm{cm}$ \citep{Kasliwal17,Gottlieb18gammarays,Matsumoto19}. There are two main potential physical scenarios to explain GRB\,170817A. GRB\,170817A might have been produced by some dissipative process within less-energetic wide-angle jet ``wings'' around the 
ultra-relativistic jet core (e.g., \citealt{Lazzati2017post170817,Meng18,Ioka19,Kathirgamaraju2019EMcounterparts,Geng19}), or by  shock breakout emission of the cocoon (inflated by the jet) 
as it emerged from the merger ejecta (\S\ref{subsec:counterparts}, \citealt{Nakar18,Gottlieb18gammarays,Bromberg18}).  
Both scenarios involve the presence of a mildly relativistic, laterally extended outflow component, yet with some key differences discussed below. We note that these two scenarios are not mutually exclusive: the cocoon shock breakout emission is a robust phenomenon for both hydrodynamic and magnetic jets, with magnetic jets being able to develop structure even after breakout as a consequence of the propagation of a rarefaction wave   \citep{Bromberg18,Geng19,Kathirgamaraju2019EMcounterparts}.

 The cocoon shock breakout scenario requires $R_{\gamma}$ to coincide with the breakout radius $R_{bo}$ from the merger ejecta, which in turns requires the existence of a fast tail of merger ejecta with $v\sim 0.6-0.8$c (much faster than the bulk of the 
 ejecta; \S\ref{Sec:Thermal}) able to reach $R_{\gamma}$ at  $\Delta t_{GW-\gamma}$ 
 (\citealt{Bromberg18,Gottlieb18gammarays}). Interestingly,  BNS simulations have suggested the presence of such a fast tail of dynamical ejecta with $v \gtrsim 0.6c$ and mass up to $\sim10^{-5}\,\rm{M_{\odot}}$ originating from the interface of the merging NSs  (e.g., \citealt{Bauswein13,Hotokezaka13,Kyutoku14,Hotokezaka18}).
  In the jet-wings scenario, $R_{\gamma}$ has instead no knowledge of the fastest merger ejecta tail and it is not bound to within their location at $\Delta t_{GW-\gamma}$. 

An interesting implication is that in the cocoon shock breakout scenario the angular time-scale $\sim R_{\gamma}/(2c\Gamma^2)$ can be too short to account for $\Delta t_{GW-\gamma}$, and $\Delta t _{inj}$ has to contribute a sizeable fraction of the observed delay (e.g., \citealt{Gottlieb18gammarays,Nakar18} and \citealt{Bromberg18} assume $\Delta t _{inj}=0.8$ s  and 0.5\,s, respectively, see also \citealt{Granot17}). In this framework, the observed $\Delta t_{GW-\gamma}\approx T_{90}$ is mostly a chance coincidence. Jet-wings scenarios can instead invoke larger angular time-scales $\sim 2$ s that regulate both the delay and the duration of the GRB \citep{Geng19,Zhang18Rate,Zhang19}: 
the observed  $\Delta t_{GW-\gamma}\approx T_{90}$ can thus be interpreted as a natural consequence of  $\Delta t_{GW-\gamma}$ being dominated by the time necessary to the jet to travel to the transparency radius, with a negligible jet launching time ($\Delta t _{inj}< \Delta t_{GW-\gamma}$). A similar situation can also happen within cocoon shock breakout scenarios, if the jet energy is near the critical minimum value for jet breakout, which leads to a small difference between the shock velocity and the homologous ejecta velocity, and hence a delayed breakout time on timescales longer than the engine duration (i.e., $\Delta t_{bo}\approx \Delta t_{GW-\gamma}$,  ``late breakout'' scenario of \citealt{Duffell18}). 
Finally, in the cocoon shock breakout scenario the peculiar light-curve morphology of GRB\,170817A and its hard-to-soft spectral evolution naturally result from the transition of the cocoon shock hydrodynamics from a planar to a spherical phase (\citealt{Gottlieb18gammarays}). In jet-wings models the spectral softening might be related to the jet structure and/or the emergence of thermal components (\citealt{Meng18}).  

To conclude, while  it is difficult to rule out either set of models with high confidence based on the $\gamma$-rays from this single event, there are  predictions that link $\Delta t_{GW-\gamma}$, $M_{ej}$, and the relative energy radiated by the $\gamma$-ray pulse and the  total afterglow kinetic energy that can be tested with a statistical sample of NS-NS (and NS-BH) mergers.

\section{\gfo: UV/Optical/IR THERMAL EMISSION} \label{Sec:Thermal} 

A worldwide effort commenced after the discovery of \gfo\ to obtain
UV, optical, and NIR photometry from many telescopes \citep{andreoni17,Arcavi17,coulter17,Cowperthwaite17,diaz17,Drout17,Evans17,hu17,Kasliwal17,lipunov17,Pian17,shappee17,Smartt17,soares17,Tanvir17,Troja17,utsumi17,valenti17,Pozanenko18}.
The combined dataset on \gfo\ contains more than 600 individual datapoints from 46 instruments (as compiled by \citealt{Villar17}).

\begin{figure}[h]
\hskip -1cm
\includegraphics[width=5.09in]{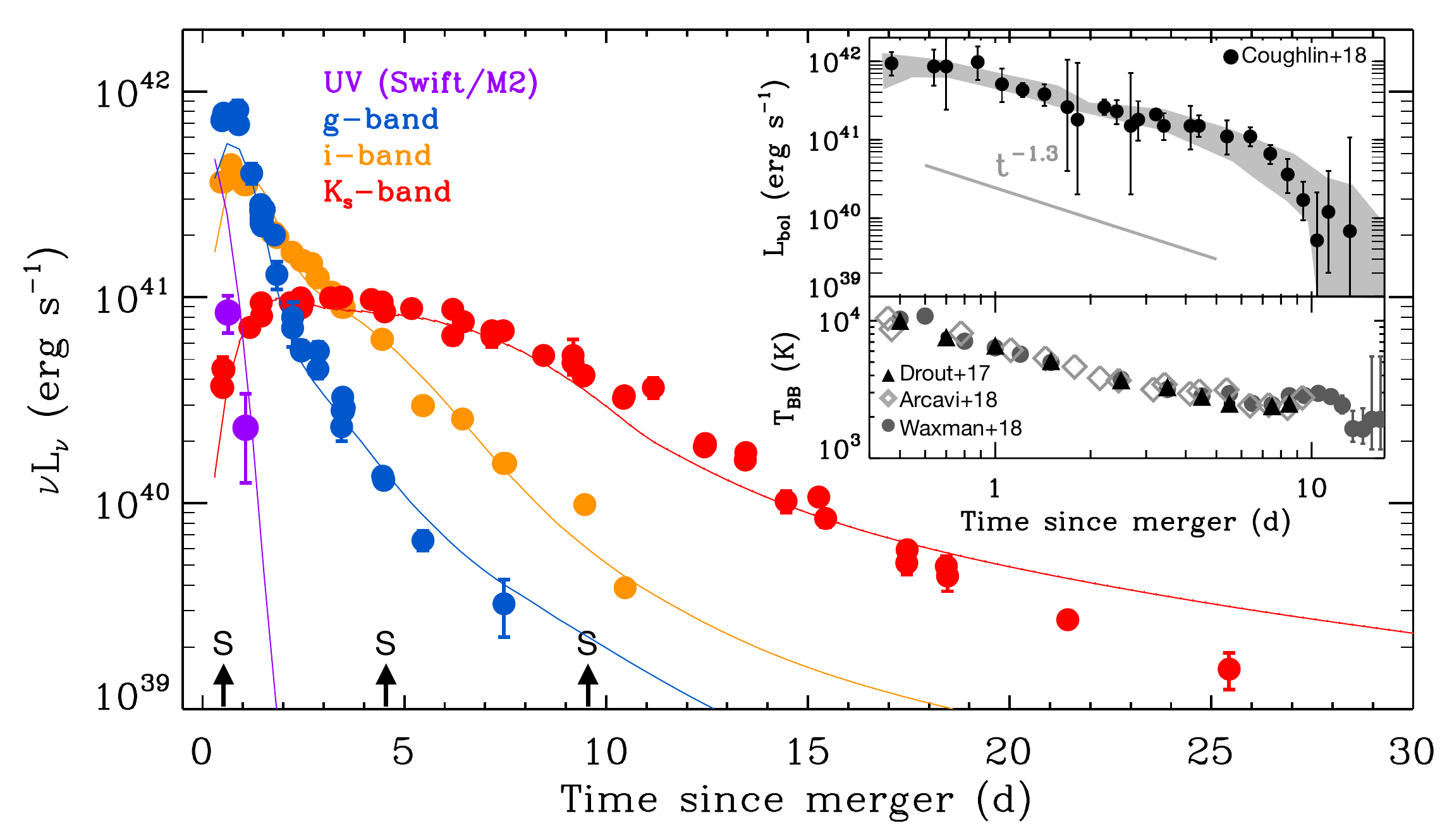}
\caption{Light curves of \gfo\ in four representative filters: \emph{Swift-}$UVM2$ ($\lambda_{\rm eff}=2231$~\AA);  $g$ ($\lambda_{\rm eff}=4671$~\AA);  $i$ ($\lambda_{\rm eff}=7458$~\AA);  $K_s$ ($\lambda_{\rm eff}=2.14~\mu$m). Data and best-fitting three-component model from 
\cite{Villar17}, with original data presented in the references cited in 
\S\ref{Sec:Thermal}.
Vertical arrows indicate the times of the spectra displayed in {\bf Figure \ref{Fig:KNSpec}}. \emph{Inset, upper panel:} Bolometric luminosity from \cite{Coughlin18} (black circles with uncertainties). The shaded gray area marks the range of best-fitting bolometric light curves from the literature \citep{Cowperthwaite17,Drout17,Arcavi18,Waxman18}. The solid gray line shows a slope of $L_{\rm bol}\propto t^{-1.3}$, the expected slope of energy input from \rp radioactive decay. \emph{Inset, lower panel:} Best fitting blackbody temperatures $T_{\rm BB}$ \citep{Drout17,Arcavi18,Waxman18}.  
Note that \citet{Drout17} fixed  $T_{\rm BB}=$2500~K after $\delta$t=8.5~d.  }
\label{Fig:KNLC}
\end{figure}

For clarity, in {\bf Figure~\ref{Fig:KNLC}} we select four representative filters with high temporal sampling to demonstrate the photometric behavior of \gfo\ from the near-UV (\emph{Swift}-$UVM2$), blue optical ($g$), red optical ($i$), through the NIR ($K_s$).  The UV light curves exhibit fading behavior from the first observations at $\delta$t=0.65~d \citep{Evans17}. At the other extreme, the $K_s$ light curve rose to a broad peak around $\delta$t$\approx$3.5~d. In between, the optical emission started fading within a day after the merger \citep{Arcavi17,coulter17,Cowperthwaite17,Kasliwal17,Pian17,Smartt17,soares17}.

Optical spectroscopy in the first week after the merger was presented by a number of groups \citep{andreoni17,Kasliwal17,Levan17,mccully17,nicholl17,Pian17,shappee17,Smartt17,Troja17,valenti17}, with the first spectrum acquired at $\delta t=0.5$~d after merger.  The spectra were unlike those of known supernovae and only developed weak features before the transient became too faint at $\delta t \approx 10$ d.
NIR spectroscopy was obtained from a few sources starting at $\delta$t=1.5~d \citep{Chornock17,Kasliwal17,Pian17,Smartt17,Tanvir17,Troja17}, and resulted in the detection of a number of broad features between 1--2~$\mu$m.

\begin{figure}[h]
\hskip -1cm
\includegraphics[width=4.8in]{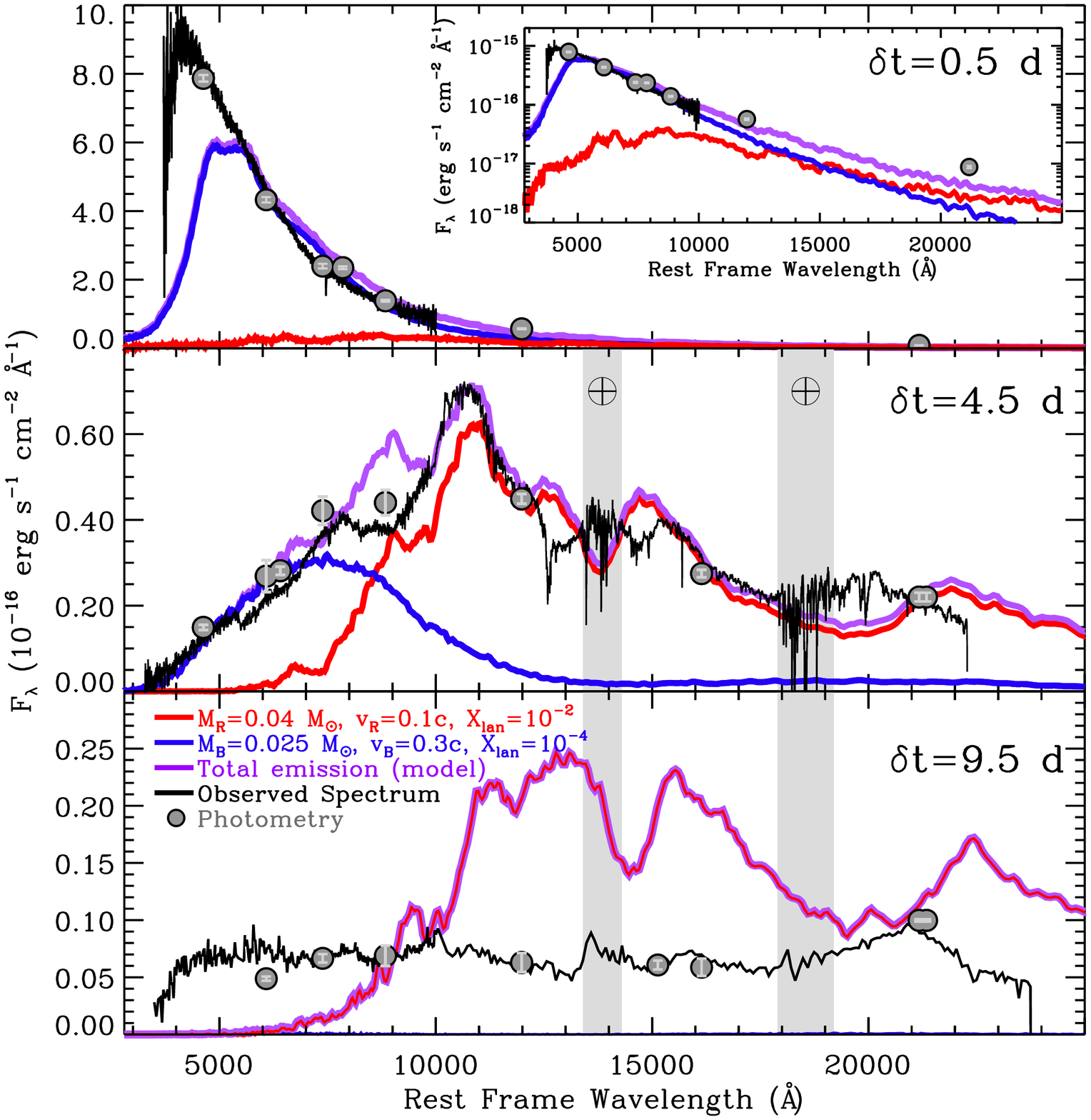}
\caption{Spectral evolution of \gfo compared to two-component kilonova models from \citet{Kasen17}.  Each panel ($\delta$t=0.5, 4.5, and 9.5~d after merger
) shows a selected spectrum in black \citep{shappee17,Pian17,Smartt17}, a low-lanthanide KN model (blue), a high-lanthanide KN model (red), and their sum (purple). The KN models have the properties listed in the legend. Gray circles show the photometry interpolated from the compilation of \citet{Villar17}. \emph{Inset:} same data and models as in the main panel at $\delta$t=0.5~d, but with a logarithmic flux scale to better show the NIR excess over the pure low-lanthanide blue KN model even at early times.
Gray bands mark the regions most affected by telluric absorption.}
\label{Fig:KNSpec}
\end{figure}

We display three epochs of spectroscopy in {\bf Figure~\ref{Fig:KNSpec}} to sample the evolution of the spectral energy distribution (SED). The first epoch ($\delta$t=0.5~d) is dominated by a blue component of emission \citep{shappee17}, while the second epoch ($\delta$t=4.5~d) was chosen to demonstrate the epoch with the clearest broad features in the NIR \citep{Chornock17,Pian17,Smartt17}. The final epoch ($\delta$t=9.5~d) shows the persistence of some of those NIR features even as the transient became difficult to observe and the signal-to-noise ratio became poor. 

\subsection{Basic properties and relationship to other optical transients} 
	
	Here we summarize the most important phenomenological features of \gfo: \\
$\bullet$ \emph{Peak bolometric luminosity of $\sim$10$^{42}$ erg~s$^{-1}$:} This is comparable to the peak luminosities of some core-collapse supernovae, and substantially greater than some of the prior predictions from kilonova models incorporating lanthanide opacities. \\
$\bullet$ \emph{Thermal SED at early times:} The SED is reasonably well characterized by a blackbody function at the earliest epochs that we have data. In particular, the first UV data at $\delta t=0.6$~d result in an SED that has too much curvature to be well fit by a power-law function, which is inconsistent with the synchrotron spectra of GRB afterglows \citep[e.g.,][]{Evans17}. Deviations from a blackbody SED grew over time. \\
$\bullet$ \emph{Fast fading in the optical:} In the seven days after the first optical detection, the transient faded by $3.90\pm0.18$~mag in $i$ and $5.87\pm0.34$~mag in $g$ \citep[\textbf{Figure~\ref{Fig:KNLC}};][]{Cowperthwaite17,Drout17,siebert17,Villar17}.  \\
$\bullet$ \emph{Large expansion velocities of the optical photosphere:} \citet{shappee17}  analyzed the first spectrum of \gfo ({\bf Figure~\ref{Fig:KNSpec}}) and found a color temperature of $T_{\rm BB}$=11,000$^{+3400}_{-900}$~K (90\% confidence levels). Given the luminosity of the source, the material at the photosphere had to be ejected at a velocity of $0.26^{+0.02}_{-0.07}c$ to reach the large implied photospheric radius (3.3$^{+0.3}_{-0.8}\times10^{14}$~cm) in the $\delta$t=0.50~d after the merger. Similar analyses based on blackbody radii inferred from either spectra or photometry require expansion velocities to be around 0.3$c$ during the first day and remain above 0.1$c$ for the first $\sim$7 days after the merger \citep{Cowperthwaite17,Drout17,Evans17,Kasliwal17,nicholl17,Pian17,Troja17}. \\
$\bullet$  \emph{Rapid cooling of the photospheric temperature:} Cooling is apparent in the photometry even within 1~d of the merger ({\bf Figure~\ref{Fig:KNLC}}). The peak of the SED moved out of the optical and longwards of 1 micron by $\delta t \approx 5$~d. After $\sim$8~d, the color temperature of the ejecta ($T_{\rm BB}$) asymptotically approached $\sim$2500\,K, a very unusual value.  \\
$\bullet$  \emph{Lack of supernova-like features in optical spectra:} The initial optical spectra were smooth and blue, with a peak near $\sim$4000~\AA\ \citep{shappee17}. Over the next few days, the spectra developed a peak that rapidly moved red towards the NIR, but never exhibited the clear P-Cygni features found in supernovae \citep{kilpatrick17,mccully17,nicholl17,Pian17,shappee17,Smartt17}. Instead, a few broader features ($\Delta \lambda$/$\lambda \approx$ 0.1--0.2) developed at red optical wavelengths (minima near 7400 and 8300~\AA) and in the NIR \citep[{\bf Figure~\ref{Fig:KNSpec}};][]{Chornock17,Pian17,Smartt17,Tanvir17,Troja17}. The severe blending of spectral features was interpreted as requiring high expansion velocities of 0.2--0.3$c$ for the material emitting in the optical and $\sim$0.1$c$ for the material that dominated the NIR emission.	\\ 

This combination of properties is unprecedented for an optical transient. No previously known supernova light curve fades as rapidly as \gfo \citep[e.g.,][]{Arcavi17,siebert17,Smartt17}. The spectra do not resemble those of known classes of transients. Two separate arguments, line blending and the required expansion velocities of the inferred blackbody radii, lead to the conclusion that the material dominating the optical emission in the first few days expanded at velocities of roughly 30\% the speed of light, high even for a supernova.

Particular focus is due the unusually red colors of \gfo after the first week ($g-K_s \gtrsim 5$~mag; $T_{\rm BB}\approx 2500$~K). 
 This is an extraordinary SED for an astronomical transient. 
 By contrast, the hydrogen-rich atmospheres of Type II supernovae asymptotically approach color temperatures around 5000~K on the photometric plateau due to the jump in opacity provided by hydrogen recombination near this temperature. The emission from the iron-rich ejecta of Type Ia supernovae is not well described by a blackbody spectrum, but the color temperatures are also near 6000~K due to the wavelength dependence of the opacity of iron-peak elements. This immediately implies that the ejecta of \gfo\ must have sources of opacity (and, presumably, a composition) unlike those found in normal supernovae. \citet{gall17} demonstrate that dust formation does not provide a plausible alternative interpretation.
 Notably, \citet{kasen13} predicted that the recombination of Nd and other lanthanide elements would occur at temperatures near 2500~K, providing a natural explanation for a kilonova to have a photospheric temperature regulated to around this value.

\subsection{Kilonova models} 
\label{subsec:KNmodels}

The light curves of astronomical transients powered by diffusion of energy deposited by radioactive decay rise to a peak that occurs when the diffusion timescale through the ejecta is comparable to the time after explosion \citep{Arnett82}. The luminosity near the peak is roughly equal to the instantaneous rate of energy deposition for many reasonable models.
In the case of an \rp-powered kilonova, \citet{metzger19kn} provides these relationships:
\begin{equation}
    \begin{aligned}
        t_{\rm peak} & \approx 1.6~{\rm d} \left( \frac{M}{10^{-2}~{\rm M}_{\odot}}\right)^{1/2} \left( \frac{v}{0.1 c} \right)^{-1/2} \left( \frac{\kappa}{1~{\rm cm}^2~{\rm g}^{-1}} \right)^{1/2} \\
    L_{\rm peak} & \approx 10^{41}~{\rm erg s}^{-1} \left( \frac{\epsilon_{\rm th}}{0.5} \right) \left( \frac{M}{10^{-2}~{\rm M}_{\odot}}\right)^{0.35} \left( \frac{v}{0.1 c} \right)^{0.65} \left( \frac{\kappa}{1~{\rm cm}^2~{\rm g}^{-1}} \right)^{-0.65}
\end{aligned}
\end{equation}
where $M$ is the mass of the ejecta, $v$ is the scale velocity of the ejecta, $\epsilon_{\rm th}$ is a parameter describing the thermalization efficiency of input radioactive heating, and $\kappa$ is a gray opacity.  
It can immediately be seen that the high peak luminosity of \gfo\ and short optical rise time are incompatible with high opacities.
A similar analysis led  \citet{Drout17} to conclude that the emitting material in the first 0.5~d is constrained to have an opacity of $\kappa < 0.08$~cm$^{2}$~g$^{-1}$.

There are a number of important simplifying assumptions underlying these expressions, especially that $\kappa$ is constant throughout the ejecta, even over time as the ejecta cool and recombine. The energy input originates from a superposition of numerous radioactive isotopes, whose half-lives are  distributed approximately uniformly in time, which results in a power-law heating rate \citep{li98}. Detailed studies have found that the overall heating rate is $\propto$$t^{-1.3}$, which we show as the gray line in the inset of {\bf Figure~\ref{Fig:KNLC}} \citep{metzger10,barnes16}, although the normalization depends on the assumed nuclear mass model \citep[e.g.,][]{rosswog17}. Another key ingredient is the thermalization efficiency, which is expected to decrease over time  \citep[e.g.,][]{kb19,hotokezaka20}. However, by assuming $\epsilon_{\rm th}\approx0.5$ at early times and taking an estimate of the expansion velocities from spectroscopy, the model reduces to a pair of equations to solve for the required opacities and ejecta masses to produce the observed timescales and luminosities.

Diffusion-powered light curve models can be fit to the bolometric light curve of a transient by specifying a gray opacity motivated by the expected composition. With a prescription for the  color temperature (e.g., from blackbody considerations), the fits can be performed on individual filters. These Arnett-like light curve models with only a single component of emission do not fit \gfo\ very well because of the difference in timescales between the blue and NIR light curves.
	Instead, the range of outflow properties seen in numerical simulations of BNS mergers motivates exploration of multi-component models. Then one component with a small diffusion timescale can dominate in the bluest filters and at the earliest times, while a longer timescale component can supplement the emission at later times and in the NIR \citep{Cowperthwaite17,Drout17,kilpatrick17}. As an example, the best fit three-component model from \citet{Villar17} is shown in {\bf Figure~\ref{Fig:KNLC}}.

	To move beyond Arnett-like models requires sophisticated radiative transfer modeling based on the latest atomic data.
	The quantitative set of inferences about composition and ejecta masses for \gfo\ in the initial observational papers in 2017 were primarily based on comparisons to only three sets of underlying radiative transfer calculations \citep{Kasen17,tanaka18,wollaeger18}.
	One important distinction is that \citet{Kasen17} and \citet{tanaka18} use an expansion opacity formalism to treat the millions of bound-bound line transitions \citep{karp77,eastmanpinto93}, while the models of \citet{wollaeger18} instead used a ``line-smeared" approach to opacities \citep{fontes17}, which tends to result in even higher lanthanide opacity. Subsequent to \gw, all groups have continued to refine the underlying atomic physics and numerical techniques \citep[e.g.,][]{kb19,fontes20,tanaka20}.
	
	Another fundamental difference in the approaches taken by these models is that both \citet{tanaka18} and \citet{wollaeger18} anchor their models to ejecta properties and composition distributions (\ye) from numerical simulations of nucleosynthesis in BNS merger ejecta, including both dynamical ejecta and winds \citep{perego14,rosswog14,wanajo14}. The models of \citet{Kasen17} have only three free parameters, a total mass of the ejecta, an average velocity (defined as $v=\sqrt{2E_{\rm k}/M}$ with $E_{\rm k}$ the kinetic energy), and a fractional lanthanide concentration \xlan. Because the open $f$-shell lanthanides dominate the opacity whenever they are present, \xlan captures the most important parameter of the underlying nucleosynthesis for affecting the kilonova SED. \citet{tanaka20} connect \xlan and $\kappa$ to different \ye ranges seen in their radiative transfer models under typical kilonova conditions. The highest opacities ($\kappa\approx$ 20--30~cm$^2$~g$^{-1}$) and lanthanide abundances ($\xlan \approx$ 0.1--0.2) are provided by material with $\ye < 0.2$. Conversely, material with $\ye \approx0.4$ has essentially zero lanthanide production, and yet still has an opacity of $\kappa\approx$ 1~cm$^2$~g$^{-1}$ due to the light \rp elements.  Ejecta with intermediate lanthanide abundances ($\xlan \approx 10^{-3}-10^{-2}$) are produced in a narrow range of \ye near 0.25, and have opacities of $\kappa \approx 5$~cm$^2$~g$^{-1}$.

	In the top panel of {\bf Figure~\ref{Fig:KNSpec}}, we compare a low-lanthanide blue kilonova model to the earliest spectrum of \gfo \citep{Kasen17,kilpatrick17,shappee17}. The velocity parameter was chosen to be 0.3$c$ to match the inferred expansion velocity of the optical photosphere and \xlan was constrained to be below $\sim$10$^{-4}$ to match the optical/NIR flux ratio. The mass of 0.025~M$_{\odot}$ is then necessary to match the flux level. By the epoch of the middle panel ($\delta$t=4.5~d), the blue kilonova had faded and cooled. The prominent NIR features are a good match for a red kilonova model with mass 0.04~M$_{\odot}$, $v=0.1c$, and a higher lanthanide abundance of \xlan=0.01  \citep{Chornock17}. \citet{Kasen17} identify Nd as the likely dominant source of these features. Taking the same kilonova models and extrapolating them forward to $\delta$t=9.5~d results in the poor fit in the bottom panel. The models overproduce the flux at this epoch, which can be somewhat alleviated by reducing the overall mass in each component, at the expense of a worse match at early times. At this late time, the radiative transfer or thermalization assumptions may be violated as the ejecta transition to being optically thin.

    Fits using the models of \citet{wollaeger18} find that the early emission from \gfo\ can be reproduced using a relatively massive (0.01--0.03~M$_{\odot}$) and moderately neutron rich ($\ye \approx 0.27$) post-merger wind component, with a subdominant neutron-rich dynamical ejecta component necessary to produce NIR emission on longer timescales \citep{Evans17,Tanvir17,Troja17}. \citet{tanaka17} found that 0.03~M$_{\odot}$ of material with a broad range of compositions (\ye = 0.1--0.4) would underproduce the early blue emission. Adding an ejecta component with moderate \ye  to improve the early light curve increases the total required ejecta mass to $\sim$0.05~M$_{\odot}$ \citep{Pian17}.
	
	One caveat to this discussion has to do with the interpretation of \xlan. Detailed nucleosynthesis calculations of NSM ejecta find that the production of lanthanides is robust once a strong \rp proceeds \citep[e.g.,][]{korobkin12,wanajo14}. The parameter study of \citet{lippuner15} finds some dependence of the final lanthanide abundance on the initial entropy and expansion timescale of a merger ejecta parcel, but the primary determinant is \ye. If it is below a threshold value of $\ye\approx0.25$, nucleosynthesis proceeds to the third peak of \rp abundances, and $\xlan\approx0.1$. For \ye somewhat above this threshold, the lanthanide abundance is negligible. Intermediate values require unrealistically narrow ranges of \ye, or macroscopic mixing of high and low lanthanide abundance material to produce an effective \xlan in the observed range.

\subsection{Evidence for \rp}
\label{subsec:rprocess}

A very elementary observation supporting the presence of \rp nucleosynthesis in \gfo is that there was an optical/NIR transient to be seen at all, particularly after the first day. The inferred blackbody radii at $\delta$t$\approx$0.5~d were $\sim$4$\times 10^{14}$~cm and grew beyond 10$^{15}$~cm in the subsequent days, while the merger ejecta were launched from within a few NS radii of the remnant. In the absence of a long-lived source of heating, the ejecta would rapidly lose their internal energy by expanding many orders of magnitude in scale. 
Even in cocoon models, the inferred radioactive heating of the transient dominates the luminosity after the first day because the cocoon also cools \citep{Kasliwal17,Duffell18,Gottlieb18cocoonUV}.  Other models for the early blue emission discussed below are also applicable only at the earliest times. Normal supernova material is heated by $^{56}$Ni, but fits of toy models powered by nickel decay would require $\sim$75\% of the ejecta to be composed of radioactive nickel \citep[e.g.,][]{Cowperthwaite17}, which is in contradiction with the observed SED and spectral features. The radioactive species formed as a by-product of \rp nucleosynthesis provide a natural match to the luminosity and timescale needed to explain \gfo, particularly for the long-lived NIR component, and any alternative model must provide a solution to this question.

The fits to the data in the previous section using several models with different assumptions all imply that \gw\ resulted in the ejection of $\sim$0.05~M$_{\odot}$ of \rp material. This needs to be combined with the BNS merger rate to determine whether it matches the required \rp\ production rate to explain Galactic nucleosynthesis, as estimated from the observed abundances of elements such as Eu.  \citet{hotokezaka15} have estimated that if a class of sources synthesizes 10$^{-2}$~M$_{\odot}$ of heavy \rp (atomic mass $A>90$) material, the required event rate (averaged over the star-formation history of the Galaxy) is only $\sim$50~Myr$^{-1}$. This can be compared with the estimated current rate for BNS mergers in the Galaxy of $\mathfrak{R}_{\rm BNS} = 37^{+24}_{-11}$~Myr$^{-1}$ (90\% confidence; \citealt{pol20}), although note that the merger rate is believed to have been higher in the past. Alternatively, the GW-derived BNS rate of $\mathfrak{R}=980^{+1490}_{-730}\,\rm{Gpc^{-3}yr^{-1}}$ \citep{Abbott20latestBNSrate} and the density of massive galaxies of $\sim$0.01~Mpc$^{-3}$ can be combined to estimate an average  rate of BNS mergers of $\sim$100~Myr$^{-1}$ for typical massive galaxies, which would require ejection of $\sim$0.005~M$_{\odot}$ of heavy \rp material per event. In either case, there is easily sufficient production of \rp material in \gw, assuming that it is typical of BNS mergers, and it is even a little bit high relative to expectations (e.g., \citealt{rosswog18}). 
Ultimately, moving beyond these order-of-magnitude estimates towards better understanding of the role of BNS mergers in Galactic nucleosynthesis will require comparison of the detailed measured abundance patterns of stars \citep[e.g.,][]{holmbeck20rpa} to numerical simulations of Galactic formation that resolve gas dynamics \citep[e.g.,][]{shen15,vdv15} as well as the accretion of disrupted satellite galaxies \citep[e.g.,][]{roederer18}.

The resolution of this 60-year-old mystery is a strong claim to make without having provided a confident identification of any particular spectral feature. Instead, the claim is based on the unusual features of the SED ($T_{\rm BB}\approx2500$~K) and broad spectral features, which reflect theoretical predictions for the unique opacity of material enriched in lanthanides. We also did not present evidence to constrain the detailed abundance pattern beyond estimates of \xlan. However, the universality of the solar \rp\ abundance pattern in metal-poor stars \citep{sneden08} gives us confidence that if lanthanides have been synthesized, then the other heavy \rp\ elements must be present as well at nearly the standard ratios, consistent with the predictions of a wide \ye distribution in BNS merger ejecta \citep{korobkin12,Bauswein13,rosswog14,wanajo14}. The unusual spectra at early times are also consistent with predictions for material composed of the light \rp \citep{banerjee20}.

The primary reason for this somewhat indirect approach (beyond line blending at these ejecta velocities) is that the radiative transfer models described in the previous section are based on atomic structure calculations that still have significant uncertainties. The resulting line lists of bound-bound transitions thus lack the accuracy necessary to be confident of the precise wavelengths of individual transitions. An alternative modeling philosophy is to start with highly accurate, but incomplete, line lists and try to identify a few of the strongest features. This approach was used by \citet{watson19}, who found that a few very broad lines ($\sim$0.2$c$) of Sr II could match the strongest deviations from a blackbody in the optical spectrum of \gfo\ over the first few days after the merger.

The multi-component kilonova ejecta picture has also been challenged by \citet{Waxman18}, who find a good fit to the bolometric light curve with a single low-opacity component corresponding to $\xlan\approx10^{-3}$, which would be insufficient production to account for the solar system \rp\ abundances. This is similar to the low opacity and $\ye=0.25$ models of \citet{Smartt17} and \citet{tanaka17}, respectively.
We note that this does require fine-tuning of the ejecta \ye because of the strong dependence of lanthanide production on this parameter \citep{lippuner15}. An alternative is that there is sufficient macroscopic mixing of the ejecta from two components with very different lanthanide abundances as to approximate the overall opacity of a single intermediate component. 
Actinide abundances in \rp-enhanced metal-poor stars may also point to the requirement for material with the lowest \ye from the dynamical ejecta to be mixed with other ejecta components with higher \ye \citep{holmbeck19actinide}. 
\citet{ji19} have also questioned whether \gw\ produced sufficient heavy \rp\ material to account for abundance ratios in metal-poor stars.
It is thus important to consider if there are any other lines of evidence that constrain the presence of the heaviest \rp isotopes in the merger ejecta (\S\ref{subsubsec:lateobs}). 

\subsubsection{Late-time IR Observations}
\label{subsubsec:lateobs}
By early September 2017, ground-based optical observations of \gfo became increasingly difficult due to its rapid fading into the bright background of its host galaxy and then impossible after it entered solar conjunction. Subsequent epochs of optical/IR photometry were obtained with the {\it Hubble Space Telescope} ({\it HST}) and the {\it Spitzer Space Telescope}. The {\it HST} observations only detected the late-time afterglow emission and are discussed in \S\ref{Sec:Afterglow}.
{\it Spitzer} was able to obtain two epochs of 3.6 and 4.5~$\mu$m photometry at $+$43 and $+$74~d after the merger \citep{villar18,kasliwal19}. \gfo is detected in both epochs at 4.5~$\mu$m, but not 3.6~$\mu$m.
The fluxes at 4.5~$\mu$m are significantly brighter than the inferred afterglow contribution at that wavelength and thus represent the latest detections of \gfo. \citet{kasliwal19} found that the steep decay between the two {\it Spitzer} epochs was indicative of a small number of heavy isotopes with half-lives around 14~d powering the radioactive transient. Potentially, this represents our best evidence for the production of second and third peak of \rp\ elements in the merger ejecta. Observations of future kilonovae with the {\it James Webb Space Telescope (JWST)} might directly detect the signatures of these heavy elements  \citep{zhu18,wu19kn}.

\subsection{Early blue emission}
\label{SubSec:BlueEmission}

One of the most unanticipated results from \gfo\ was the luminosity of the early blue emission.
The short risetime also places severe constraints on the opacity of the emitting material (\S \ref{subsec:KNmodels}).
The representative blue kilonova model shown in the top panel of {\bf Figure~\ref{Fig:KNSpec}} matches the luminosity at that epoch, but it experiences too much line blanketing in the blue and underproduces the IR emission (which is remedied by the contribution from the red component).
One of the major uncertainties for modeling kilonovae at these early times comes from the paucity of appropriate atomic data for the unusual ejecta conditions. \citet{banerjee20} have recently produced opacity calculations for the relevant highly ionized heavy elements and find that they can approximately reproduce the early light curve of \gfo with $M\approx 0.05$~M$_{\odot}$ of lanthanide-poor light \rp material.

The unexpected properties of \gfo in the first day after the merger have motivated an exploration for alternative models to produce extra blue emission beyond that expected purely from radioactively heated ejecta.
Cocoon models produce two effects that can increase the early optical luminosity  \citep{Kasliwal17,Gottlieb18cocoonUV}. The first is the direct cooling of the jet-deposited energy, which only contributes over the first few hours. The second is the Doppler-boosted emission from the cocoon material, which is moving at mildly relativistic velocities.

Another model was proposed by \citet{Piro18}, who argued that the asymmetric light curve of \gfo\ was not typical of objects whose light curves are governed on both the rise and fall by the same diffusion timescale from a central energy source (e.g., radioactive decay) and instead proposed that merger ejecta surrounding the remnant at larger radius were shock heated and subsequently radiated. However, this picture was challenged by the hydrodynamic simulations of \citet{Duffell18}, who noted the importance of the fact that merger ejecta would be expanding homologously rather than stationary. Their numerical calculations 
concluded that for collimated jets with $E_k < 10^{51}$ erg, shock heating due to the jet propagation into the BNS ejecta is energetically subdominant (contributing $10^{48}-10^{49}$ erg on time scales of 0.1--1~s) and represents a minor contribution to the luminosity of thermal optical transients on longer timescales.

\subsection{Relationship to components in BNS merger simulations}

We are now in a position to relate the phenomenological components inferred from the optical observations on \gfo to the various mass-ejection components in BNS merger simulations.
The total inferred ejecta mass is too high for only the tidal dynamical ejecta \citep{Sekiguchi16}, which are also too neutron rich to produce the early optical emission from a lanthanide-poor component \citep{korobkin12}. 
The shock-heated dynamical ejecta can reach sufficiently high velocities to be consistent with the early optical spectra and can be lanthanide poor \citep{wanajo14}, but the ejecta masses from this mechanism appear to be too low ($\lesssim0.01$~M$_{\odot}$) unless the NS radius is very small \citep{oechslin07,Bauswein13,Hotokezaka13}. 
Winds from the accretion disk can produce outflows of a range of compositions, but the velocities may not be sufficiently high to match the early observations \citep{kasen15winds,fahlman18}.
However, post-merger winds provide the most natural explanation for the high ejecta mass and material being present with a wide range of \xlan \citep{grossman14,metzger14bluekn,rosswog14,just15}.
The relative importance of winds driven by neutrinos \citep{dessart09,metzger14bluekn,perego14}, magnetic fields \citep{Metzger18magnetar}, and viscous effects \citep{Radice18BNS} remains an open question.

The numerical simulations of BNS mergers result in highly aspherical ejecta, with high-opacity material commonly being produced in the equatorial plane. The luminosity and SED of the resulting emission is likely to be dependent on the viewing angle, with the general trend being higher luminosities, particularly in the blue, resulting from a more polar viewing angle \citep[e.g.,][]{wollaeger18}.
Despite this effect, many of the studies inferring ejecta parameters from the light curves or spectra of \gfo\ have used effectively one-dimensional radiative transfer models.  In the case of the multi-component models discussed in \S \ref{subsec:KNmodels}, the flux of the separate components was simply summed to produce the total emission. This can be justified if the ejecta components have separate spatial distributions (e.g., polar vs. equatorial) and there is no radiative coupling. 
However, \citet{kawaguchi20} have identified several effects in their two-dimensional models, including photons diffusing preferentially in the direction of low opacity and heating of tidal dynamical ejecta by emission from the post-merger ejecta, that can combine to reduce the inferred total ejecta mass by as much as a factor of $\sim$2 compared to one-dimensional estimates.
There is a clear need for further development of multidimensional radiative transfer models for kilonovae to gain precision in estimates of the \rp\ ejecta mass \citep{perego17,kawaguchi18,wollaeger18,kawaguchi20,korobkin20}.

\subsubsection{Optical Polarization}
\label{sec:pol}
While the kilonova ejecta are too distant to spatially resolve the various components, the polarization of light can be a useful tool to constrain the geometry. The optical polarization signatures in explosive transients are produced in a competition between the linearly polarizing effects of electron scattering and the depolarizing effects of bound-bound line transitions. If a distant, spatially unresolved source is circularly symmetric when projected on the plane of the sky, the angles of polarization produced locally within the ejecta cancel when integrated over the photosphere. Measurable polarization is thus a signature of deviations from sphericity, and has been well studied in the case of supernovae \citep{wangwheeler08}. The highly aspherical geometries of BNS merger ejecta provide a promising avenue to generate polarization. 

\citet{covino17} were able to obtain five epochs of optical polarimetry of AT2017gfo, four in $R$ band and one in $z$. Only the first epoch, at $\delta t=1.46$~d, had a measurable polarization of $0.50\pm0.07$\%, while the others had upper limits consistent with this value. Several foreground stars had polarization measurements of similar magnitude and position angle, indicating that most of the observed polarization was produced by propagation through the interstellar medium of the Galaxy. \citet{bulla19} concluded that the intrinsic polarization of light from \gfo itself was $<0.18$\% at the 95\% confidence level. Their models showed that future polarimetric observations to probe kilonova ejecta geometry would be most informative at early times, when the low-opacity blue emission is dominant. However, the very high lanthanide line opacities will suppress any polarization signal after emission from the red component dominates the optical light.

\subsection{Comparison to kilonovae in SGRBs} \label{SubSec:KNeSGRBs}
\begin{figure}[h]
\hskip -1cm
\includegraphics[width=3.4in]{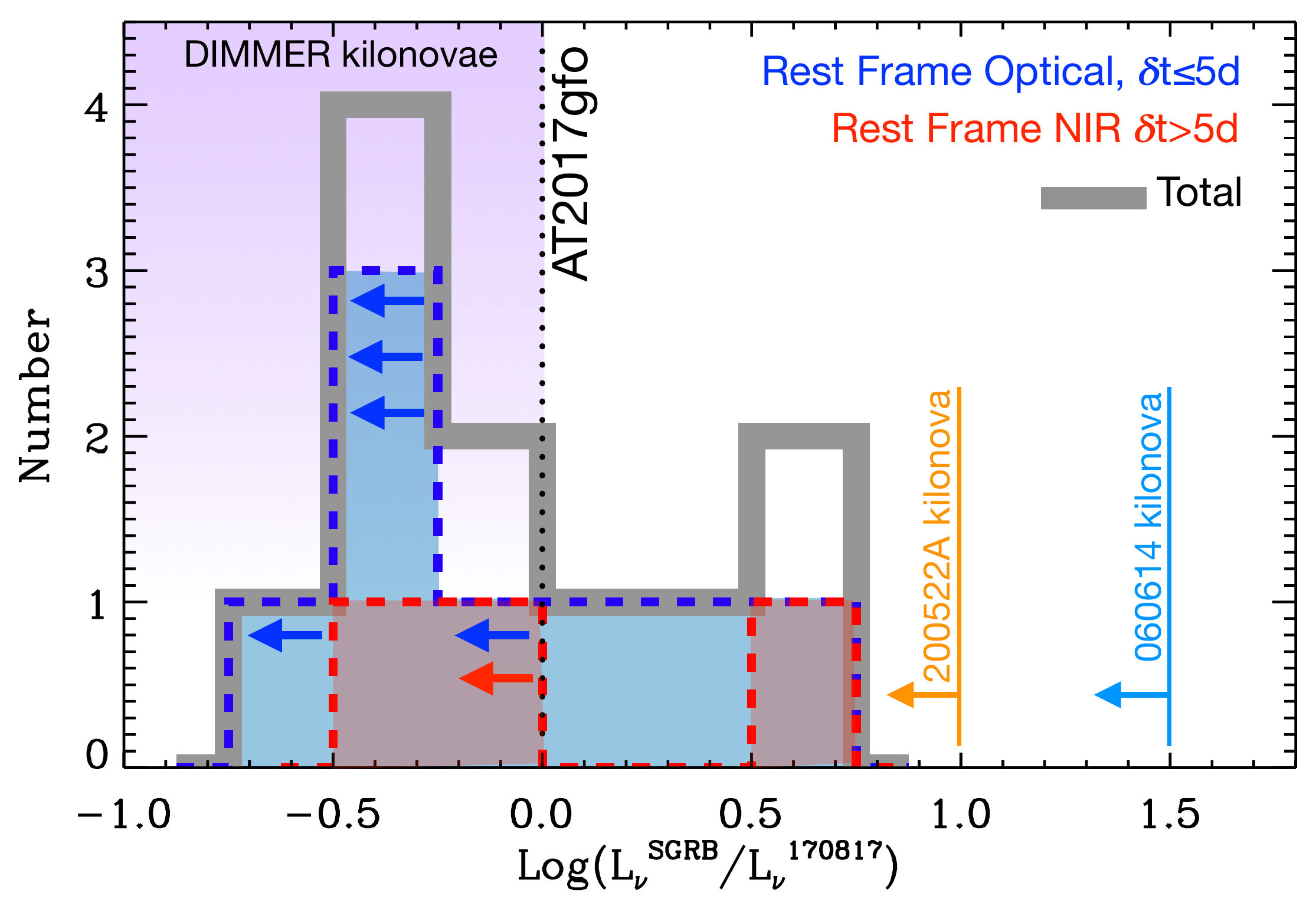}
\caption{Luminosity distribution of optical counterparts to SGRBs 
with $z\le0.5$ compared to AT\,2017gfo, highlighting the diversity of the potential kilonova emission.   Only data from bona-fide SGRBs for which a kilonova candidate has been identified or SGRBs with brightness limits that are constraining with respect to AT\,2017gfo  are shown.  Rest-frame NIR data (red) have been collected at $\delta t>5$ d with the exception of SGRB\,160624A. Optical data (blue) have been collected at earlier times $\delta t\le5$ d. Kilonovae in SGRBs like 130603B can be significantly more luminous than AT\,2017gfo, while existing upper limits 
rule out an AT\,2017gfo-like transient, pointing at the existence of fainter kilonovae or, alternatively, at a class of SGRBs not accompanied by \emph{any} kilonova emission (e.g., NS-BH mergers would allow that possibility). Optical and NIR limits on kilonovae in SGRBs 060614 and 200522A are shown in light-blue and orange, respectively. Data 
from \cite{Gompertz18diversityKN,Fong200522A,Rossi20}. 
}
\label{Fig:SGRBKNe}
\end{figure}	

Studies of nearby SGRBs ($z\lesssim 0.5$) 
have brought to light a significant diversity in the optical emission following SGRBs, which in some cases can be attributed to kilonova emission above the level of the optical afterglow.  
There are six SGRBs with potential kilonova emission detected, with different levels of observational evidence: SGRBs 
050709 \citep{Jin16}, 
060614 \citep{Yang15}, 
070809 \citep{Jin20}, 
130603B \citep{Berger13,Tanvir13}, 
150101B \citep{Gompertz18diversityKN, Troja150101B}, and 
160821B \citep{Kasliwal160821B,Lamb160821B,Troja160821B}. The recently detected SGRB\,200522A might provide the first example of a magnetar-boosted kilonova \citep{Fong200522A}.
In addition, SGRBs 
050509B, 
061201,
080905A and 160624A have deep limits that rule out an \gfo-like kilonova \citep{Gompertz18diversityKN,Ascenzi19,Fong200522A,Rossi20}.

A direct comparison to \gfo is challenging due to the very sparse nature of SGRB data (both in terms of spectral and temporal coverage) and a level of contamination by the SGRB afterglow that is difficult to quantify in most cases. The combination of these factors  makes it virtually impossible to map the observed diversity of the emission of SGRB kilonova candidates into a constrained physical parameter space of ejecta masses, ejecta velocity, and opacity. The important conclusion is that the current sample of observations of SGRB kilonovae supports the existence of a broad range of kilonova luminosities ($\approx$0.3--10 times the luminosity of AT\,2017gfo depending on the epoch and frequency of observation), \textbf{Figure \ref{Fig:SGRBKNe}}. Note that the presence of a successful SGRB oriented towards our line of sight implies that we are viewing these kilonovae from a nearly polar direction, which is believed to be the most luminous viewing angle and least affected by a possible equatorial structure with high opacity. From a population perspective, the complementary (and unsuccessful except for \gw) search for kilonovae from GW-detected NS mergers also leaves open the possibility that a large fraction of kilonovae from BNS mergers are intrinsically fainter than AT\,2017gfo \citep{Kasliwal20}.


\section{GW170817: NON-THERMAL  EMISSION} 
\label{Sec:Afterglow}
\begin{textbox}[b]\section{Jet Terminology}
\emph{Structured Jet}: Generic term for an anisotropic outflow with angular and/or radial structure and a core of ultra-relativistic material of angular size $\theta_{jet}$.  Structured jets have angle-dependent bulk Lorentz factor $\Gamma(\theta)$ and energy per unit solid angle $d E(\theta)/ d \Omega$ extending to $\theta_{w}$ with $\theta_{w}$$>$$\theta_{jet}$.  \emph{Off-axis structured jet}: structured jet for which $\theta_{obs}>\theta_{jet}$ (not necessarily $\theta_{obs}>\theta_{w}$).
\emph{Quasi-spherical outflow}: Uncollimated outflow with potential radial structure and mild angular structure. \emph{Cocoon:} Wide-angle mildly relativistic outflow created by the interaction of the relativistic jet with the merger ejecta.  A successful jet+cocoon system is a physical manifestation of a structured jet, while pure cocoon models belong to the quasi-spherical outflow category.
\end{textbox}

\begin{figure}[h]
\hskip -1. cm
\includegraphics[scale=0.65]{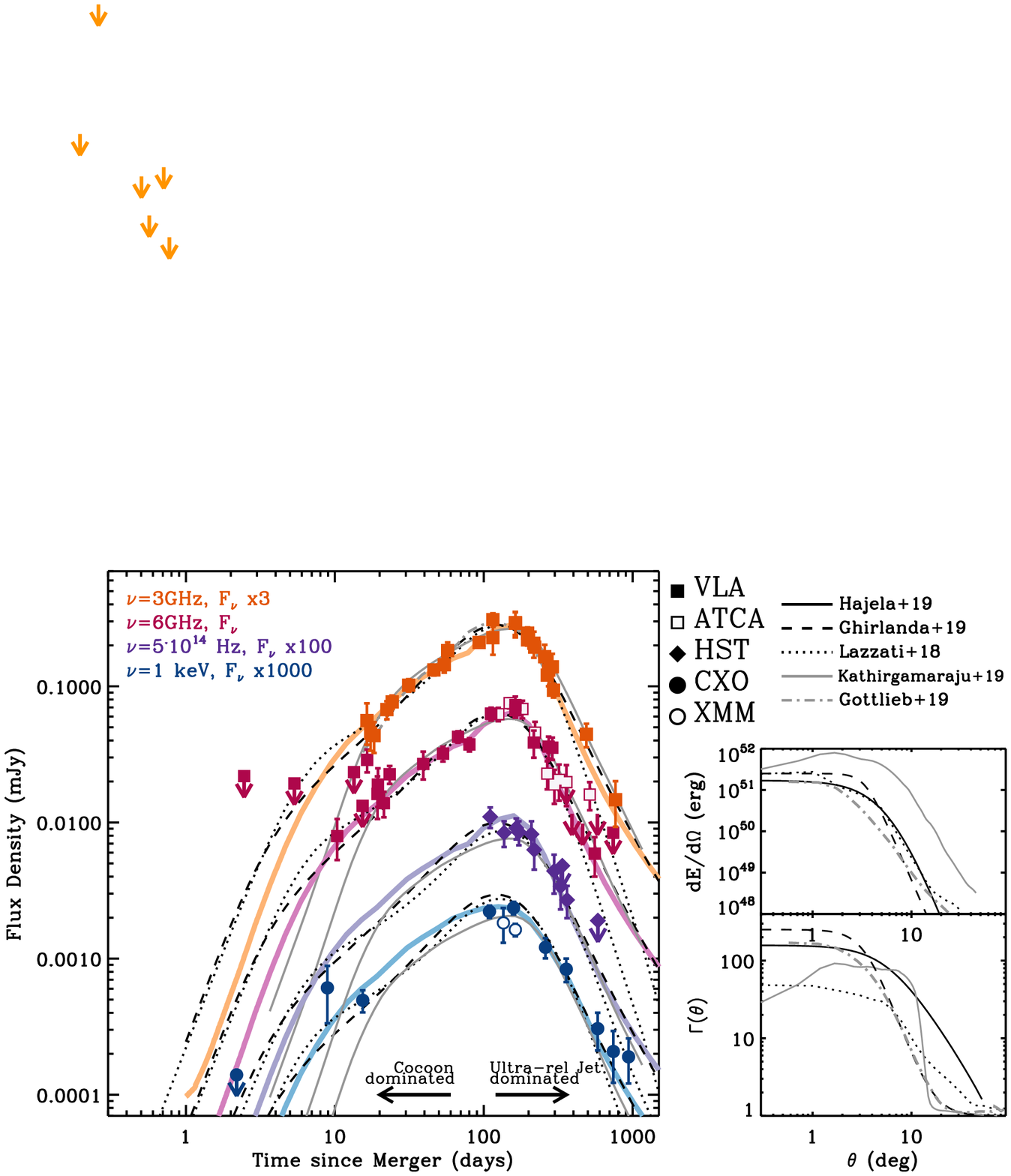}
\caption{ 
Jetted outflows with different angular structures $\Gamma(\theta)$ and $dE(\theta)/d\Omega$ (right panels) successfully reproduce the broad-band afterglow observations of GW\,170817 at radio (orange and pink), optical (purple) and X-ray (blue) wavelengths. Some models are motivated by the physics of BNS mergers  \citep{Gottlieb18gammarays,Kathirgamaraju2019EMcounterparts,Hajela19,Lazzati18}   and others are analytical abstractions (e.g., gaussian jets, \citealt{Ghirlanda19}). 
These models share the presence of a highly collimated core of ultra-relativistic ejecta 
at $\theta<\theta_{jet}$ viewed off-axis ($\theta_{obs}>\theta_{jet}$) and surrounded by mildly relativistic wings of material. 
Due to relativistic beaming, the pre-peak emission 
is dominated by radiation from the wider-angle mildly relativistic outflow (e.g., a cocoon). The jet core dominated the detected emission at $t\gtrsim t_{pk}$. Observational data originally presented by: \cite{Alexander17,Alexander18GW,Haggard17,Hallinan17,Margutti17GW,Margutti18GW,Kim17,Troja17,Troja18peak,Troja18RiseFall,Troja20,Dobie18,Lyman18,DAvanzo18,Mooley18Jet,Mooley18superluminal,Nynka18,Resmi18,Ruan18,Fong19,Hajela19,Lamb19,Piro19, Makhathini2020}.}
\label{Fig:afterglow}
\end{figure}

 Mass outflows 
 from BNS mergers drive shocks that radiate broadband synchrotron emission. This process  converts the shocks' kinetic energy into radiation, with a peak of emission that intrinsically occurs on the outflow's deceleration timescale.  
 For relativistic jets seen off-axis, the time of the observed peak of emission further depends on the geometry of the system (i.e., $\theta_{obs}$ and $\theta_{jet}$). Two conclusions follow: (i) lighter ejecta components (e.g., relativistic jets; \S\ref{SubSec:JetAfterglow}) produce synchrotron emission with an intrinsically earlier peak than that associated with more massive outflows from the merger (e.g., the kilonova ejecta;  \S\ref{SubSec:KNafterglow}); (ii) synchrotron emission is a probe of both the energy and geometry of the outflows and of the density of matter surrounding the binary at the time of merger (\S\ref{SubSec:JetAfterglow}), which is ultimately responsible for the deceleration of the mass outflows. 

Following the SGRB literature, we refer to this non-thermal emission as  an afterglow. 
The synchrotron emission depends on the outflow kinetic energy $E_{k}$, the environment density $n$, the fraction of post-shock energy into tangled magnetic fields $\epsilon_B$ and accelerated electrons $\epsilon_e$, as well as on the details of the distribution of non-thermal relativistic electrons 
$N(\gamma_e)\propto \gamma_e^{-p}$ 
(e.g., \citealt{Sari98}). In the case of collimated relativistic outflows, the observed emission carries further dependencies on  
$\theta_{jet}$  and 
$\theta_{obs}$ (\textbf{Figure \ref{Fig:cartoon}}).
As of $\sim3$ yrs after the merger, the non-thermal emission from  \gw\, has been dominated by the afterglow of 
a structured jet seen off-axis (\S\ref{SubSec:JetAfterglow}, \textbf{Figure \ref{Fig:afterglow}}). Future observations of this very nearby system  
might identify the first kilonova afterglow (\S\ref{SubSec:KNafterglow}).

\begin{figure}[h]
\includegraphics[scale=0.35]{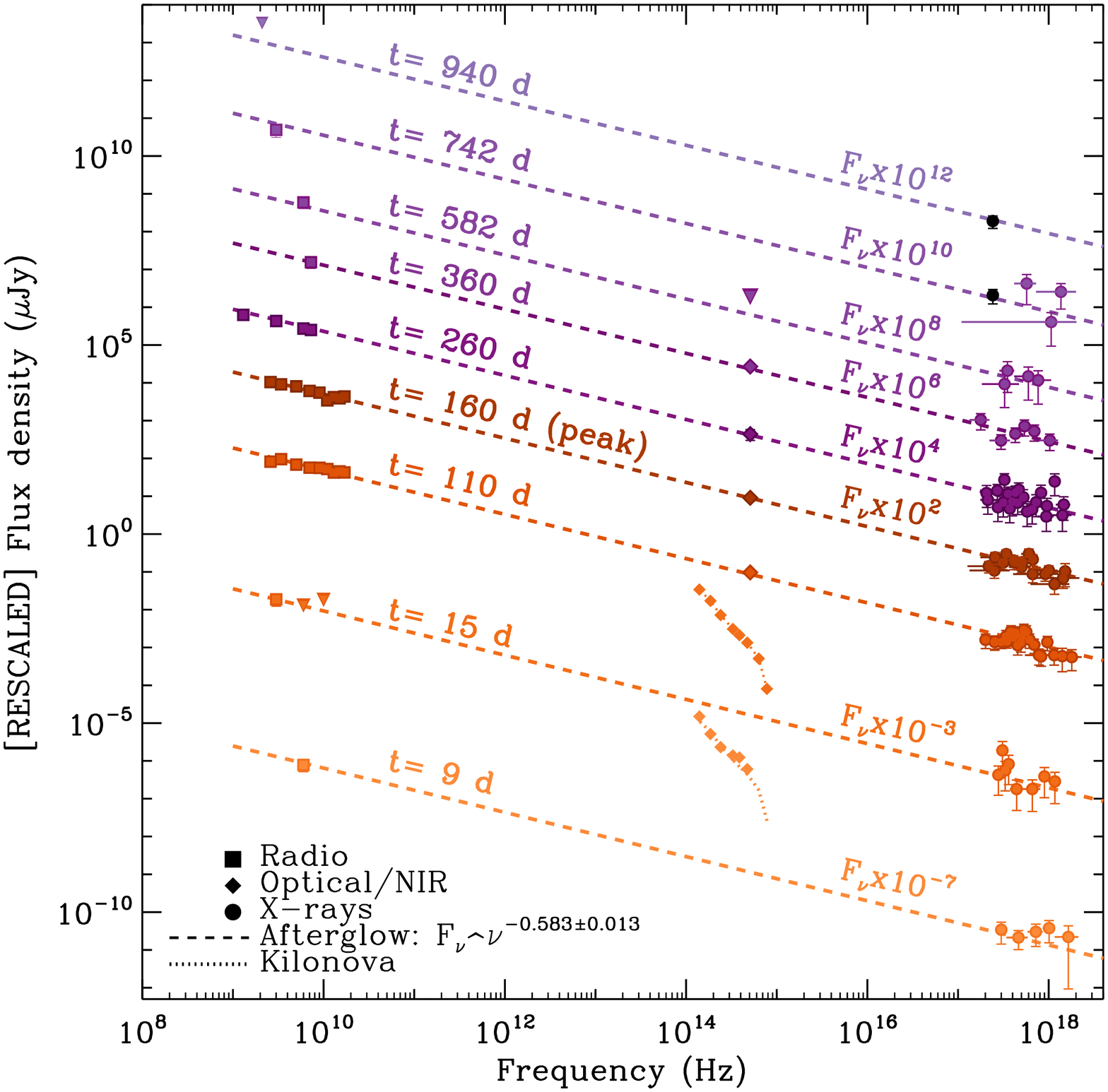}
\caption{Broad-band SED evolution of the jet afterglow of GW\,170817. X-ray (circles) and radio (squares) frequencies are dominated by non-thermal synchrotron emission on a simple power-law spectrum $F_{\nu}\propto \nu^{-\beta_{XR}}$ with $\beta_{XR}=\betaradioX$ at all times (dashed lines). During the first weeks the optical  bands (diamonds) are dominated by thermal kilonova emission  (dotted lines, \S\ref{Sec:Thermal}). 
Black points: 1-keV flux density with flux calibration performed \emph{assuming} the best-fitting $F_{\nu}\propto \nu^{-\beta_{XR}}$ spectrum, which is necessary with limited photon statistics. Data from references of \textbf{Figure \ref{Fig:afterglow}}.}
\label{Fig:SEDafterglow}
\end{figure}

\subsection{Structure and Geometry of a Jetted Relativistic Outflow}\label{SubSec:JetAfterglow}
Broadband afterglow observations of \gw{} 
provide the first direct evidence that BNS mergers are able to launch highly collimated relativistic jets that can survive the interaction with the local  merger ejecta, as first theorized by \cite{Paczynski86,Eichler89},  and are likely collimated by this very same process. These observations establish the first direct connection between canonical SGRBs and mergers of NSs, and offer the first view of a SGRB-like relativistic jet ``from the side'' (i.e., off-axis).

Key observations 
include: (i) deep X-ray and radio non-detections at early times $t\lesssim2$ d, which set GW\,170817 phenomenologically apart from all SGRB afterglows; 
(ii) gradual monotonic rise of the light-curve with 
$F_{\nu}\propto t^{\alpharise}$  
at  $10\le \delta t \le 150$ d (\textbf{Figure \ref{Fig:afterglow}}); (iii) sharp achromatic light-curve peak  at $\delta t\sim$\tpeak; (iv) steep post-peak decay  $F_{\nu}\propto t^{-\alphadecay}$ at $\delta t\ge300$ d; (v) a radio-to-X-ray spectrum that is well described at all times by a  power-law model $F_{\nu}\propto \nu^{-\beta}$ with $\beta$$\sim$$\betaradioX$ (\textbf{Figure \ref{Fig:SEDafterglow}}); (vi) superluminal motion of the centroid of the unresolved radio image of the blast wave.   
Below we describe how this combined observational evidence leads to one concordant physical scenario of a  
highly collimated ($\theta_{jet}\approx$\,\thetajet) ultra-relativistic jet directed 
away from our line of sight and carrying  $\lesssim$\Ekjet that developed wide-angle mildly relativistic ($\Gamma$$\sim$ few)  wings as it propagated through the sub-relativistic merger ejecta (i.e., an off-axis structured relativistic jet).

The extremely well-behaved  power-law spectrum $F_{\nu}\propto \nu^{-\betaradioX}$  (\textbf{Figure \ref{Fig:SEDafterglow}}) 
implies that radio and X-ray radiation (and optical as well at $\delta t$$>$$100$ d) are part of the same optically thin synchrotron spectrum, 
where the cooling frequency $\nu_c$ is above the X-ray band  and the synchrotron frequency $\nu_m$ is below the radio band at all times. In this regime $F_{\nu}\propto \nu^{-(p-1)/2}$, which leads to the most precise measurement to date of the index $p$ of the relativistic electrons distribution  
$N(\gamma_e)\propto \gamma_e^{-p}$ accelerated by a BNS merger shock: $p=$\pelectron (\citealt{Fong19}; see also \citealt{Hajela19,Lamb19,Makhathini2020,Troja20}). 

\begin{figure}[t]
\includegraphics[scale=0.58]{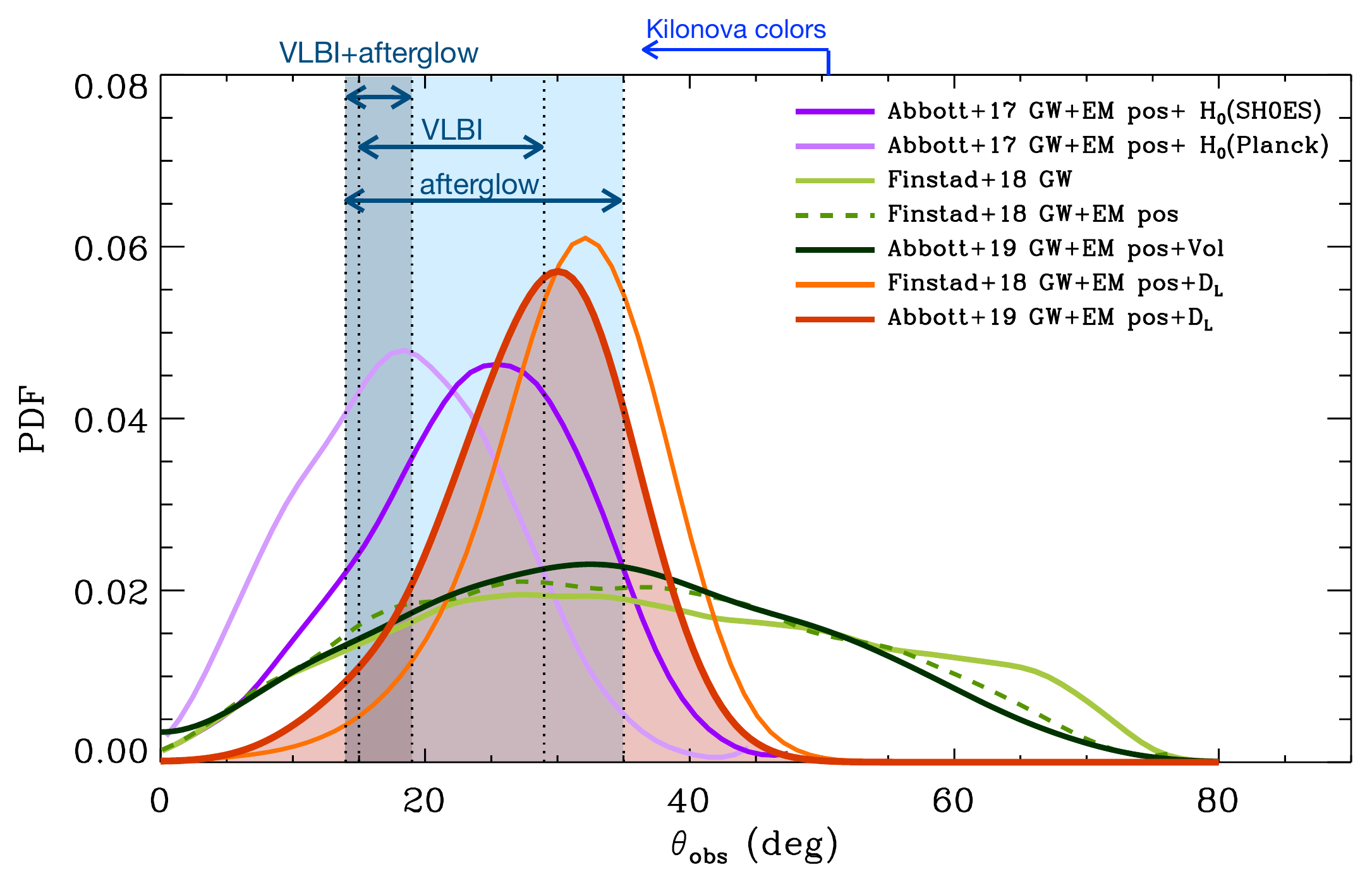}
\caption{Constraints on $\theta_{obs}$ from GWs (light-green line) combined with different priors. ``EM pos'': sub-arcsec sky localization from the EM counterpart; ``$D_{L}$'': gaussian prior on the luminosity distance inferred 
by \cite{cantiello18}; ``$H_{0}$'': Planck  or SH0ES  $H_0$ value assumed, from \cite{PlanckH0,Riess16H0}, respectively; ``Vol'': 
flat prior in volume \citep{Abbott17H0,Finstad18,abbott19properties}. Knowledge  of the precise sky location has limited effect, 
while a prior on $D_{L}$ (or $H_{0}$) strongly influences the inference on $\theta_{obs}$ (see also \citealt{Mandel18,Chen19}). The difference between the \cite{abbott19properties} and the \cite{Finstad18} GW+EM pos+$D_{\rm{L}}$ posterior is related to slightly different choices of priors and frequency of the signal. 
Shaded blue areas: best-fitting ranges for $\theta_{obs}$ from the VLBI and afterglow modeling by Ghirlanda19,Hotokezaka19, from the VLBI-driven inferences of \cite{Mooley18superluminal}, and for a variety of afterglow models that reproduce observations extending to $\delta t>1$ yr (\S\ref{Sec:Afterglow}). }
\label{Fig:ThetaObs}
\end{figure}

VLBI observations 
provided evidence for an apparent superluminal motion of the radio source centroid with an  average velocity $v_{app}=\beta_{app}c=(4.1\pm 0.5)c$ between 75--230 d \citep{Mooley18superluminal} and constrained the apparent size of the unresolved radio source  to $<2.5$ milliarcseconds at 270.4 d (90\% c.l., \citealt{Ghirlanda19}). Taken together these measurements rule out 
quasi-spherical outflows (e.g., \citealt{GillGranot2018,Granot18images,Zrake18} for simulations) and point at a compact radio source originating from a highly anisotropic outflow with average  $\Gamma\approx \beta_{app}\approx 4$ around the time of afterglow peak. 
In the limit of a relativistically moving point source these observations also provided an estimate of the geometry of the dominant source of emission $(\theta_{obs}-\theta_{jet})\sim 1/\Gamma \approx 0.25$.   
Supporting  this scenario is the sharp light-curve peak at $t_{peak}$$\sim$\tpeak followed by a steep achromatic $F_{\nu}\propto t^{-\alphadecay}$ afterglow decay, which is naturally explained as the signature of the core of a narrow relativistic jet entering our line of sight and dominating the detected emission at $t\gtrsim t_{peak}$ (\textbf{Figure \ref{Fig:afterglow}}).
Radio observation at 244~d also 
indicate a degree of linear polarization   $\Pi<12$\% (99\% c.l., frequency of 2.8 GHz), which suggests that the post-shock magnetic field cannot be fully contained within the shock plane \citep{Corsi18,GillGranot2020}.

\begin{figure}[h]
\includegraphics[scale=0.5]{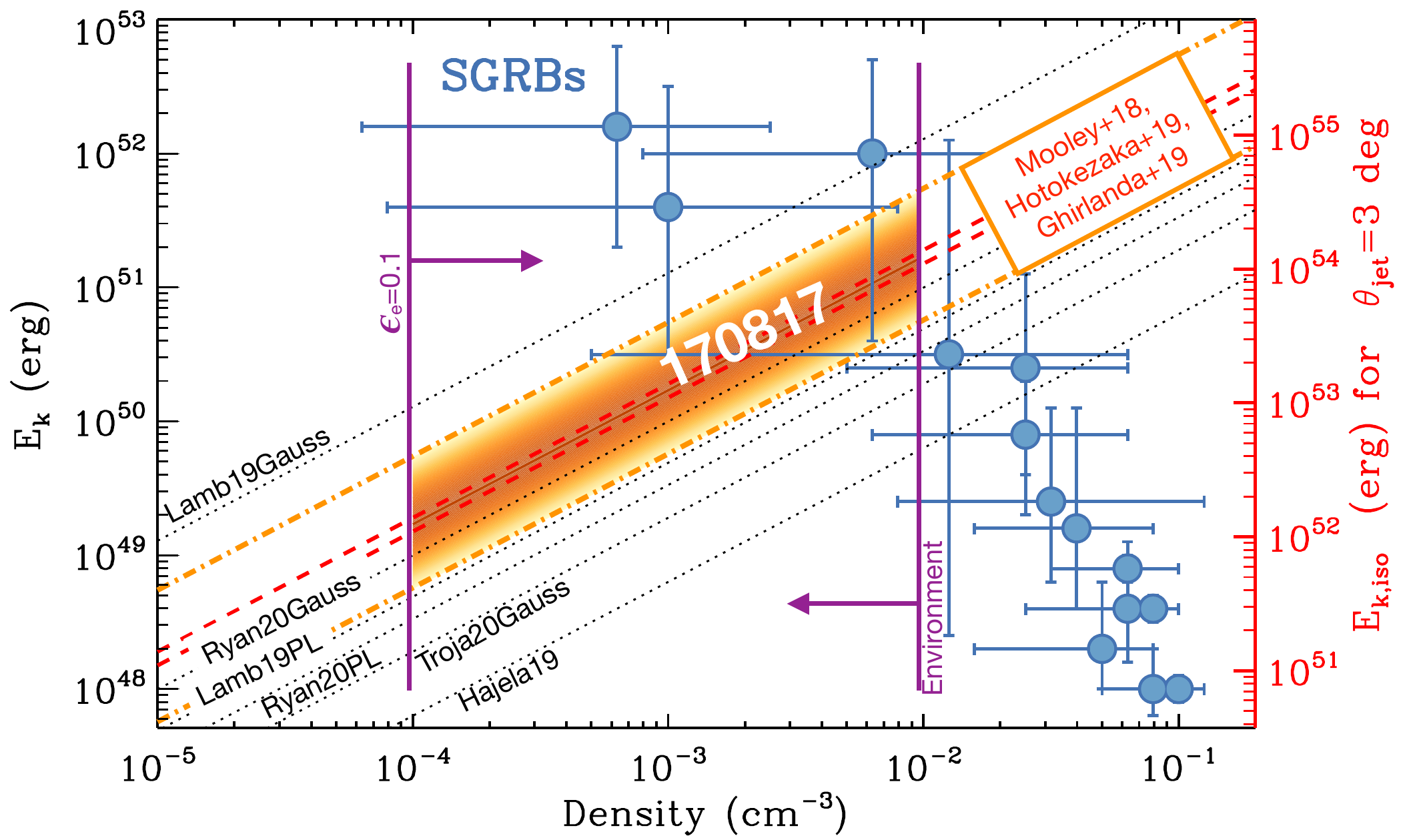}
\caption{Constraints on the jet kinetic energy (beaming-corrected and isotropic equivalent) and  environment density of \gw{} in the context of SGRBs. The red dashed lines and the orange dashed-dotted lines mark the best-fitting values and their representative  uncertanties from studies that self-consistently model the VLBI data and the afterglow data \citep{Mooley18superluminal,Hotokezaka19,Ghirlanda19}. Black dotted lines: best fitting parameters from \cite{Hajela19,Lamb19,Ryan20}  and \cite{Troja20} (for power-law and Gaussian jets) that model the long-term evolution of the \gw{} afterglow under different assumptions. Blue circles: SGRBs 
from the homogeneous  
multi-wavelength afterglow modeling  by \cite{Wu19}, where no assumption is made on $\epsilon_e,\epsilon_B$. Colored area: likely $E_k-n$ range for \gw{} based on additional constraints from host-galaxy observations (\S\ref{Sec:environment}) and theoretical expectations from the physics of particle acceleration in relativistic shocks (\S\ref{SubSec:JetAfterglow}). }
\label{Fig:Ekdensity}
\end{figure}

This steep post-peak decay is consistent with the universal post-jet-break expectation from relativistic jets, $F_{\nu}\propto t^{-p}$,  
 and contains no information on the jet collimation \citep{Lamb18postpeakdecay}. However, the rapid transition from peak to the asymptotic power-law decay on timescale $\Delta t/t_{peak}\approx$1--2  implies $\theta_{obs}/\theta_{jet}\approx$\thetaobsthetajetratio
 (e.g., \citealt{NakarPiran20, Ryan20}). When combined with the VLBI constraints, \citealt{Mooley18superluminal,Ghirlanda19,Hotokezaka19}
 find $\theta_{jet}\approx$\thetajet and $\theta_{obs}\approx$\thetaobs (see also \citealt{GillGranot2018}). The inferred $\theta_{obs}$ is consistent with inferences from the kilonova colors (\S\ref{Sec:Thermal}) and GW modeling (\S\ref{Sec:GWs}), \textbf{Figure \ref{Fig:ThetaObs}}. 
Similarly, being determined by the hydrodynamics of the deceleration of the jet core within the environment,  
$t_{peak}\propto (E_{k}/n)^{1/3} (\theta_{obs}-\theta_{jet})^{8/3} $ 
constrains the system parameters as  $(E_{k}/n)\approx$\Ekdensityratio \citep{Mooley18superluminal, Hotokezaka19, Ghirlanda19} when using the VLBI information (\textbf{Figure \ref{Fig:Ekdensity}}). Importantly, the inferences so far do not depend on the poorly known shock microphysical parameters $\epsilon_e$ and  $\epsilon_B$.

Next we consider the  physical implications of the early afterglow evolution. The deep radio and X-ray non-detections at $\delta t \lesssim 2$ d (\textbf{Figure \ref{Fig:afterglow}} and \textbf{\ref{Fig:afterglowcomparison}}) imply that the observer's line of sight is misaligned with respect to the jet core, i.e., $\theta_{obs}>\theta_{jet}$. However, the subsequent mild rise $F_{\nu}\propto t^{\alpharise}$ (\textbf{Figure \ref{Fig:afterglow}}) 
significantly less steep than $F_{\nu}\propto t^3$, indicates that the observer \emph{was} within the cone of emission of \emph{some} outflow material with angular extent $\theta_{w}$ \citep{NakarPiran18,Ryan20} 
since the time of the first afterglow detection at $\sim$9 d \citep{Troja17,Hallinan17}.
Additionally, the rising afterglow emission before peak is sensitive to the ratio $\theta_{obs}/\theta_{jet}$ \citep{Granot18offaxislightcurve,Ryan20,NakarPiran20} and requires an increase of \emph{observed} energy per unit time $E_{obs}\propto t^{1.3}$ \citep{NakarPiran18, Pooley18},  which can be either the result of true energy injection into the shock (e.g., due to the deceleration of a radially stratified isotropic fireball with $\Gamma(r)$), or due to increasing energy per unit time that intercepts the observer's line of sight (i.e., apparent energy injection due to  the progressive decrease of relativistic beaming of an anisotropic outflow with $\Gamma(\theta)$ and $E(\theta)$). 

The combined evidence from the steep post-peak decay and VLBI observations rules out quasi-spherical radially stratified fireballs and points at an outflow with some angular structure, i.e., a structured jet. However, due to a massive degeneracy between $E(\theta)$, $\theta_{obs}$ and $\Gamma(\theta,t)$, the pre-peak afterglow light-curve does \emph{not} provide a unique $E(\theta)$ solution, in spite of being a direct manifestation of structure in the outflow \citep{NakarPiran20,Ryan20,BeniaminiGranot20}. For Gaussian jets with $E(\theta)\propto e^{-(\theta/\theta_{jet})^{2}}$ the observed rise implies $\theta_{obs}/\theta_{jet}\sim$\thetaobsthetajetratio, consistent with the findings above \citep{Ryan20}. As a result, a variety of jet angular structures can adequately fit the afterglow data (\textbf{Figure \ref{Fig:afterglow}}). 
These models are tuned to reproduce the afterglow data and are not necessarily sensitive to the tail of wider-angle mildly relativistic material that produced GRB\,170817A (\S\ref{Sec:GammaRays}; \citealt{Lamb18jetcounterpart,Ioka19}). 

We conclude with a critical assessment of the insight offered by the afterglow data \citep{Gill19,Ryan20,NakarPiran20}. 
Detailed observations of the non-thermal emission in GW\,170817 provide uncontroversial evidence for the presence of an energetic, highly collimated relativistic jet with angular structure directed away from our line of sight and provide measurements of $\theta_{obs}$, $\theta_{jet}$, and $p$.
Yet, these data leave the specific angular structure of the outflow at $\theta>\theta_{obs}$ largely unconstrained. In particular, the angular extent of the jet ``wings'' $\theta_w$ is unknown. Additionally, even after a jet structure is \emph{assumed}, the afterglow data effectively provide fewer constraints than the model parameters (i.e.,  the light curve, spectrum, and VLBI observations provide five constraints to the jet-core model vs. at least seven model parameters: $n$,  $E_{k}$, $\theta_{jet}$, $\theta_{obs}$, $\epsilon_e$, $\epsilon_B$, $p$), which partially stems from  the fact that the synchrotron spectral breaks ($\nu_{sa},\nu_{m}, \nu_{c}$) of the jet afterglow of GW\,170817A fell outside the observed spectral window (e.g., \cite{Granot18offaxislightcurve,Gill19}).
A notable consequence is that while the ratio $E_{k}/n$ is well constrained, as it effectively controls the outflow dynamics, the two parameters are not individually as well constrained by the jet afterglow modeling alone. 
Adding the independent inference on the environment density $n< 0.01\,\rm{cm^{-3}}$  (\S\ref{SubSec:HG}) leads to $E_{k}<$\Ekjet, 
independent from the shock microphysics values. $\epsilon_e$ and $\epsilon_B$ are loosely constrained by the requirement $\nu_c(t_{pk})>\nu_X$, which leads to $n\lesssim 2\times 10^{-6}\epsilon_B^{-1}\,\rm{cm^{-3}}$ and by the observed flux at peak $F_{\nu,pk}(\nu)$, which links $n$, $\epsilon_B$ and $\epsilon_e$. Taking $n\sim10^{-3}\,\rm{cm^{-3}}$ as a fiducial density value in early-type galaxies, the relations above suggest small $\epsilon_B<10^{-3}$. Specifically, for $\epsilon_e=0.1$ (a robust prediction from simulations of particle acceleration by relativistic shocks, e.g., \citealt{Sironi13}),  $n\gtrsim 10^{-4}\,\rm{cm^{-3}}$ and  $\epsilon_B \lesssim 4\times 10^{-4}$ (\textbf{Figure \ref{Fig:Ekdensity}}).

Additionally, since the transition from the coasting phase to the deceleration phase of the jet core was \emph{not} observed, the initial jet-core Lorentz factor $\Gamma_{0}$ after break out is fundamentally unconstrained (from VLBI observations, $\Gamma_{0}>4$).  A further constraint on the system will be provided by the transition to the non-relativistic regime (\S\ref{SubSec:othersources}) that leads to an achromatic flattening of the afterglow light curve as the outflow enters the Sedov phase, becomes spherical, and the emission from the counter-jet enters our line of sight.   However, other components of emission might outshine the jet afterglow by this time (\S\ref{SubSec:KNafterglow}).

\subsection{A physically motivated structured-jet model}\label{SubSec:JetAfterglowImplications}

Structured relativistic outflows are a natural outcome of BNS mergers (\S\ref{subsec:counterparts}) and the outflow's structure (radial and/or angular) and collimation can be imparted by the jet acceleration process (e.g., \citealt{Kathirgamaraju2019EMcounterparts} in the context of MHD-driven jets) or by the hydrodynamical interaction of the jet with the merger's debris cloud consisting of winds and dynamical ejecta (references in \S\ref{subsec:counterparts},  \citealt{Kasliwal17,Lazzati2017post170817,Murguia-Berthier17,Duffell18,Gottlieb18gammarays,Lazzati18,Nakar18,Xie18,Lazzati19}) or both (\citealt{Bromberg18}). 
The presence of heavy dynamical ejecta along the rotation axis, and the subsequent jet interaction, 
might have played a primary role in the jet collimation process in GW\,170817, which has been a long-standing theoretical problem in BNS mergers \citep{Nagakura14,Duffell15}. In this context, other key parameters determining the angular structure of the outflow and how the energy is partitioned within the outflow are the delay (if any) between the the BNS merger and jet launching, as well as the time the engine remains active after jet breakout (e.g., \citealt{Murguia-Berthier17,Murguia-Berthier20,Geng19,Lazzati19,Beniamini20}). The structure of the fastest outflows from BNS mergers thus encodes information about the NS EoS, the nature of the remnant (\emph{if} jet launching requires the collapse to BH), and the jet launching mechanism.  

Recent simulations of BNS mergers listed above have shown that the process involving a light jet trying to pierce through the dense merger ejecta leads to two potential outcomes: the jet is stalled within the ejecta  and no collimated ultra-relativistic outflow survives; alternatively, more energetic jets or jets that encounter less  mass enveloping the polar regions  can survive the interaction and break through the merger ejecta, launching a powerful jets in the circum-merger environment. 
For both successful and failed jets,  the propagation of the jet within the merger ejecta creates a mildly relativistic ($\Gamma<10$) wide-angle cocoon with energy $E_{c}$ proportional to the time spent by the jet within the ejecta (e.g., \citealt{Ramirez-Ruiz02}). 
The jet energy $E_{k}$ is proportional to the time the engine remains active after the jet breaks out from the merger ejecta.  In the case of failed jets, the resulting outflow consists of a pure cocoon (i.e., a wide angle $\theta\gtrsim 30^\circ$ mildly relativistic outflow). In the case of successful jets, the resulting outflow consists of a jet+cocoon system, with a narrow ($\theta_{jet}\sim$ a few degrees) highly relativistic core surrounded by a sheath of mildly relativistic material at $\theta>\theta_{jet}$. 
Since the quenching (or survival) of the jet is not theoretically guaranteed, both options are equally viable until observational evidence 
contradicts the expectations from either class of models.   

The afterglow observations of GW\,170817 (\S\ref{SubSec:JetAfterglow}) establish that some BNS mergers are able to launch relativistic jets  
that survive the interaction with the merger debris cloud (ejecta and winds)  while also powering wide-angle outflows with energy $E_c$ comparable to the jet energy $E_{k}$. 
While different jet structures 
can adequately explain the afterglow data (\textbf{Figure \ref{Fig:afterglow}}), the jet+cocoon system, with 
a built-in mechanism for dissipation of energy into $\gamma$-rays, also self-consistently accounts for GRB\,170817A (\S\ref{Sec:GammaRays}), and thus constitutes a natural physical model for GW\,170187.
Conversely, the observed non-thermal afterglow of GW\,170817 is \emph{not} consistent with quasi-spherical models including magnetar-like giant flares \citep{Salafia18}, the interaction of the fast tail of the dynamical ejecta with the environment \citep{Hotokezaka18}, or pure cocoon systems (i.e., a failed-jet scenario) that were viable options until the emission from jet core entered our line of sight at $\sim$\tpeak (e.g., \citealt{Kasliwal17,Mooley18superluminal, Nakar18, NakarPiran18}).

\subsection{Connection to SGRB afterglows} \label{SubSec:SGRBafterglows}

\begin{figure}[h]
\hskip 0. cm
\includegraphics[width=6.3in]{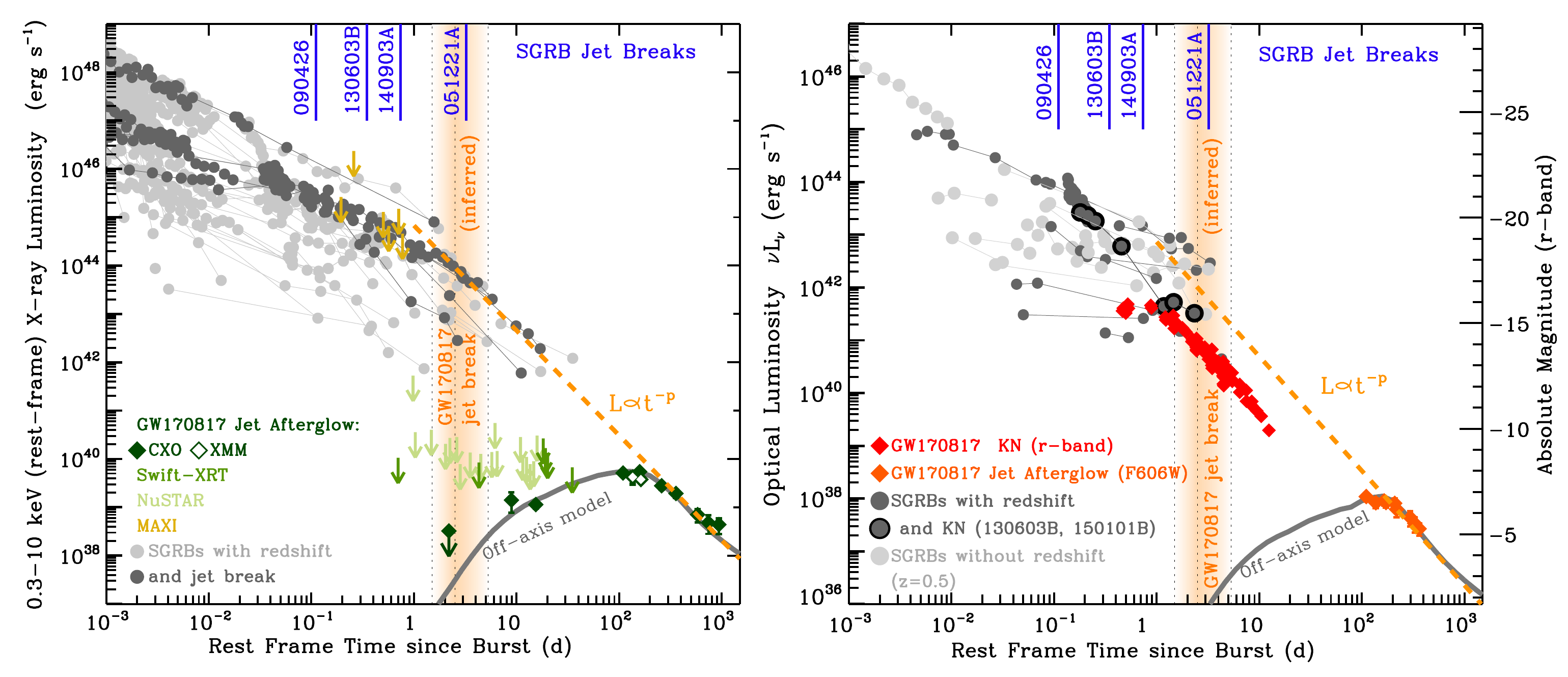}
\vskip 0 cm
\caption{Optical (right panel) and X-ray (left panel) GW\,170817 jet afterglow in the context of SGRB afterglows. Known SGRB jet-break times are indicated in blue, while the orange-shaded area marks the range of inferred jet-break times were GW\,170817 to have been observed on-axis. Interestingly, the back extrapolation of the post-peak, post-jet-break $L\propto t^{-p}$ evolution of GW\,170817 (dashed orange line) intercepts the luminosity of SGRB afterglows around the time of the inferred jet break, indicating that similar jet and environment properties are shared by GW\,170817 and SGRBs. 
The optical panel shows that the kilonova (KN) of GW\,170817 (red diamonds) typically would be outshined by the optical afterglow of an on-axis jet. The early \emph{Swift}-XRT and NuSTAR limits acquired at $t<$2 d with $L_x<10^{40}\,\rm{erg\,s^{-1}}$ clearly set GW\,170817 apart from cosmological SGRB afterglows. References: SGRBs: \cite{Burrows051221A,Grupe050724,Evans09,Fong14-GRB130603B,Fong15,Fong17,Troja140903A,Troja150101B}; \gw{}: \cite{Evans17,Sugita18} and references of \textbf{Figure \ref{Fig:afterglow}}.   } 
\label{Fig:afterglowcomparison}
\end{figure}

Observations of SGRB afterglows in the last decade provide a remarkable basis for comparison with GW\,170817, which represents the first bona-fide detection of a broadband afterglow from an off-axis ultra-relativistic jet launched by  a BNS merger. 
SGRBs harbor relativistic jets with similar energy ($E_k\approx 10^{48}-10^{52}\,\rm{erg}$) viewed on-axis (with the likely exception of SGRB 150101B; \citealt{Troja150101B}) and propagating into similarly low-density environments ($n$$\approx$$ 10^{-4}-10^{-1}\,\rm{cm^{-3}}$;  \citealt{Berger14,Fong15,Wu19}). 

Following \cite{Fong19}, we first proceed with a simple, yet model-agnostic, exercise. We place the afterglow 
of GW\,170817 in the luminosity phase space of SGRBs in \textbf{Figure \ref{Fig:afterglowcomparison}}. A few considerations follow: (i) with detections extending to 940 d since merger, the proximity of GW\,170817 is allowing us to explore a phase of the afterglow evolution that we have never sampled before; (ii) optical kilonovae are typically outshined by on-axis optical afterglows. The overlap between AT2017gfo and the faintest optical afterglows 
allows for the possibility that some SGRBs are viewed slightly off-axis \citep{Fong19}, similarly to SGRB\,150101B \citep{BurnsGRB150101B,Troja150101B}. 
(iii) While we know the radio evolution of \gw{} in striking detail, we only have sparse radio light-curves of 9 SGRBs after $\sim$$15$ yrs of investigations \citep{Fong200522A}.  
Common to all the spectral wavelengths is the faintness of the off-axis afterglows relative to the sensitivity of current instrumentation: the \gw{} afterglow at peak would only be detectable to within $\lesssim 160$ Mpc by the most sensitive X-ray and radio observatories (e.g., \citealt{Gottlieb19detectability}). 

Second, we extrapolate back in time the post-peak afterglow evolution  $F_{\nu}\propto t^{-2.2}$, which is entirely dominated by the jet-core component 
(\S\ref{SubSec:JetAfterglow}). 
This behavior is consistent with the $F_{\nu}\propto t^{-p}$ expectation from the post-jet-break dynamics of a jet with sideways expansion, which predicts universal afterglow light-curves  that depend on the true jet energy (rather than on its isotropic equivalent value) and carries no dependency on the system geometry (i.e., $\theta_{jet}$ and $\theta_{obs}$, e.g., \citealt{Granot18offaxislightcurve}).
\textbf{Figure \ref{Fig:afterglowcomparison}} shows that  the extrapolation of the post-peak evolution of the jet afterglow of GW\,170817 does intersect with the SGRB afterglow population at their expected/measured jet-break times. This is consistent with the notion that GW\,170817 and cosmological SGRBs share similar combinations of true jet-core energetics, circumburst density, and shock microphysics. Variations of these extrinsic and intrinsic properties contribute to the diversity of SGRB afterglows post jet-break, while viewing angle effects are  primarily responsible for the  \emph{observed} differences in the early-time broadband temporal evolution of SGRBs and the afterglow of GW\,170817.  

This comparison suggests that jets that successfully pierce through the merger debris might have similar ultra-relativistic \emph{core} energetics (see \citealt{Salafia19,Wu19} for a detailed calculation of the on-axis afterglow).
However, no conclusion can be drawn on the \emph{universality} of the angular structure of the outflow, because in SGRBs the emission from the jet core likely dominates at all times (with no detectable contribution from the ``wings''). The universality of the jet structure is an important open question that directly connects the merger conditions and outcome (i.e., if and when a jet is launched, the amount and distribution of ejecta and wind material along the jet path; \S\ref{SubSec:JetAfterglowImplications}), which depend on the NS EoS, and the accretion physics, which sets the accretion-to-jet energy conversion efficiency, and hence the jet energy reservoir  (e.g., \citealt{Salafia20}).  

We end with considerations on the jet collimation of \gw compared to SGRBs and on the fraction of successful jets in BNS mergers. SGRB jet opening angles measured from afterglow jet-breaks are in the range $\theta_{jet}^{SGRB}\sim 3\degree-8\degree$ \citep{Fong15,Troja160821B,Lamb160821B}. The $\theta_{jet}^{SGRB}$ measurements are biased against less collimated jets that would  break at later times and fainter fluxes not covered by observations. A few SGRBs have jet-break times lower limits indicative of wider jets with $\theta_{jet}^{SGRB}\gtrsim 13\degree-25\degree$ (e.g., SGRBs 050709, 050724A, and 120804A). In any case, with $\theta_{jet}\approx$\thetajet, \gw{} lies in the highly collimated end of the SGRB jet-angle distribution. These properties, together with current GW constraints on BNS mergers, indicate that a fraction $>$10\% of BNS mergers launch collimated jets that successfully pierce through the merger debris cloud \citep{Ghirlanda19}, with this fraction potentially extending to $\approx$100\%   (\textbf{Figure \ref{Fig:Rates}}, see also \citealt{Beniamini19}). However, the combination of the high level of collimation and intrinsic or apparent 
faintness of the wide-angle and off-axis $\gamma$-ray emission, respectively, and sensitivity of the current instrumentation implies that only $1\%-10\%$ of GW discovered BNS mergers 
will have a detected $\gamma$-ray counterpart \citep{Beniamini19}.

\subsection{Other potential sources of non-thermal emission at $t<1000$ days}\label{SubSec:othersources}

Radiation from a long-lived central engine such as an accreting BH 
or a millisecond magnetar 
has been invoked in SGRBs to power their extended emission
(e.g., \citealt{Metzger08EE}), X-ray flares \citep{Margutti11}, X-ray light-curve plateaus, 
as well as 
the late-time excess of X-rays of SGRB\,130603B 
(\citealt{Fong14-GRB130603B,Kisaka16}). 
The X-ray optical depth through the merger ejecta of density $\rho_{ej}$, mass $M_{\rm{ej}}$, radius $R_{\rm{ej}}\sim v_{\rm{ej}} t$ is
$\tau_{\rm X} \simeq \rho_{\rm{ej}} R_{\rm{ej}} \kappa_{\rm X}\approx 2\times 10^4 (\kappa_{\rm X}/1000\rm \,cm^{2}g^{-1}) (M_{\rm ej}/10^{-2}M_{\odot}) (v_{\rm ej}/0.2\rm \, c)^{-2} (t/1 d)^{-2}$, where $\kappa_{\rm X}\sim 1000$ cm$^{2}$ g$^{-1}$ is the bound-free opacity of neutral or singly ionized heavy $r$-process nuclei at $\sim$$1-10$ keV (e.g., \citealt{metzger19kn}).  
For centrally produced X-rays to be able to leak out, either $\tau_{\rm X} $$<$$ 1$, or the ejecta material needs to be fully ionized, which requires  $L_x$$\sim 10^{43}-10^{44}\,\rm{erg\,s^{-1}}$ (e.g., \citealt{Metzger14}). Since neither of these conditions are met by \gw at $t<150$ d (e.g., \citealt{Pooley18}), it is unlikely that radiation from the merger remnant dominates the X-ray energy release at those early epochs. The remarkably constant ratio of the X-ray to radio luminosity of \gw from 9 d to $\sim$750 d since merger (\textbf{Figure \ref{Fig:SEDafterglow}}) and the lack of statistically significant X-ray variability \citep{Hajela19} independently argue against a magnetar/BH origin of the detected X-rays at all times. 

Searches for short-term X-ray flux variability in future BNS mergers have the potential to uncover sudden re-activations of the central engine (\citealt{Piro19}).

\section{MULTIMESSENGER GW + EM INFERENCES} 
\subsection{Deformability, ejection of matter, and the nature of the colliding stars }
\label{SubSec:deformability}
The tidal deformability parameters $\Lambda_1$ and $\Lambda_2$ of the two compact objects describe the amount of deformation of matter. The tidal field of the binary companion induces a mass-quadrupole moment in a NS with observable effects in the GW emission that become more prominent as the orbital separation approaches the NS radius $R$. The dimensionless $\Lambda$ parameters are proportional to the induced quadrupole moment and quantify the strength of these effects. Since $\Lambda$ directly depends on the mass and radius of a NS, a measurement of $\Lambda$ directly probes and constrains the EoS of nuclear matter, as well as the intrinsic nature of the merging compact objects, as $\Lambda=0$ for a BH. Inferences on the deformability of matter can be derived from both GWs and EM emission. We first review the constraints on $\Lambda_1$ and $\Lambda_2$ placed by GW measurements alone, and then discuss the additional inferences enabled by  the detection of EM radiation from this system.

\begin{marginnote}[]
\entry{Tidal Deformability parameter}{$\Lambda\equiv \frac{2}{3}k_2\big (\frac{c^2}{G} \frac{R}{m}\big) ^5$, where $k_2$ is the dimensionless $\ell=2$ Love number, $R$ and $m$ are the NS radius and mass. 
$\Lambda=0$ for a BH as $k_2=0$ in that case.}
\end{marginnote}
\begin{marginnote}[]
\entry{Mass-weighted deformability parameter}{$\tilde \Lambda\equiv$ $\frac{16}{13}\frac{(m_1+12m_2)m_1^4\Lambda_1}{(m_1+m_2)^5}+$ $\frac{16}{13}\frac{(m_1+12m_2)m_2^4\Lambda_2}{(m_1+m_2)^5}$.  
Leading tidal contribution to the GW phase evolution defined such that  
$\tilde \Lambda=\Lambda_1=\Lambda_2$ for $m_1=m_2$. $\tilde \Lambda=0$ for a BH-BH merger. 
}
\end{marginnote}

Assuming no correlation between $\Lambda_1$ and $\Lambda_2$ and allowing the two parameters to vary independently, \cite{abbott19properties} derive a system  mass-weighted deformability parameter of $\tilde \Lambda\in (0,630)$ (90\% lower and upper limit range) for the large-spin prior ($\chi\le 0.89$), and $\tilde \Lambda=300^{+420}_{-230}$ (90\% highest posterior density interval) for the low-spin prior ($\chi \le 0.05$) that is consistent with known Galactic NSs that would merge within a Hubble time. This result 
rules out at the 90\% confidence level several EoS models (\citealt{abbott19properties}, their Fig. 10). 
The $\tilde \Lambda$ posterior has some support at $\tilde \Lambda=0$ primarily associated with binaries with low mass ratio $q=m_2/m_1\lesssim 0.5$. Interestingly, for $q\approx 1$, which is typical of the known population of Galactic BNSs, $\tilde \Lambda=0$ lies outside the 90\% confidence region of the low-spin and high-spin $\tilde \Lambda$ posteriors and the two posteriors are very similar. Only for the low-spin prior the posterior of $\tilde \Lambda=0$ has no support within the 90\% credibility interval for any $q$ value. The individual posteriors of $\Lambda_1$ and $\Lambda_2$  have support at $\Lambda_1$$=$$\Lambda_2$$=$$0$ at 90\% credibility level, for both the high-spin and low-spin priors. The conclusion is that with minimal assumptions on the intrinsic nature of the merging objects built into the priors, and based on GWs alone,  it is not possible to definitely assert that both colliding objects are NSs \citep{abbott19properties,abbott20model}: 
a BNS merger is the most likely scenario for \gw{} but 
BH-BH and NS-BH systems are statistically 
allowed.

A reasonable assumption is that both objects obey the same (unknown) EoS, which is equivalent to assuming that $\Lambda_1$ and $\Lambda_2$ are in fact correlated and have similar values for similar NS masses.  Working under this hypothesis and further assuming that the two NSs have spins in the range of those of Galactic BNSs leads to a substantial improvement on the $\Lambda_1-\Lambda_2$ credibility region (\citealt{abbott18eos}, their Fig. 1). For a NS with mass of 1.4$\,\rm{M_{\odot}}$  $\Lambda_{1.4}=190^{+390}_{-120}$ at the 90\% level, favoring ``soft'' against ``stiff'' EoSs. With the addition of EoS-insensitive relations among macroscopic properties of NSs that allow to map tidal deformabilities into NS radii, \cite{abbott18eos}  derive a primary (heavier) NS areal radius of $R_1=10.8^{+2.0}_{-1.7}\,\rm{km}$ and  $R_2=10.7^{+2.1}_{-1.5}\,\rm{km}$ for the lighter NS. Further enforcing the EoS to support a maximum NS mass 
$\geq 1.97\,\rm{M_{\odot}}$ to match the mass of the heaviest NS known ($M$$=$$2.01\pm 0.04\,\rm{M_{\odot}}$, \citealt{Antoniadis13}) the radii constraints improve to $R=11.9^{+1.4}_{-1.4}\,\rm{km}$ for both components, consistent with the results by \cite{De18,De18Erratum}.

The multiple EM counterparts of \gw{} and their associated outflows (jet, kilonova, \S\ref{Sec:Thermal}-\S\ref{Sec:Afterglow}) are a direct manifestation of the presence and ejection of matter, which ultimately implies that \emph{at least one} compact object is a NS. Several approaches have been used in the literature to combine the evidence from GWs and EM radiation into inferences on $\tilde \Lambda$, the NS EoS,  NS radii and the maximum mass of a cold spherical non-rotating NS ($M_{TOV}$, i.e., the TOV limit). Considerations  on the total energetics of the EM outflows (kilonova+jet) of the order of $\approx$$ 10^{51}$ erg (\S\ref{Sec:Thermal}-\S\ref{Sec:Afterglow}), which pose a direct constraint on the budget of extractable energy from the rapidly rotating merger remnant, together with the constraints on the kilonova ejecta masses, the presence of the blue kilonova component and the system total mass derived from GWs have been employed by \cite{MargalitMetzger17,Shibata17,Rezzolla18,Ruiz18} to derive credibility intervals of $M_{TOV}$, found to be in the range $2.01-2.33\,\rm{M_{\odot}}$. 
Relaxing the assumption of an initially rapidly rotating remnant leads to  $M_{TOV}\lesssim 2.3\,\rm{M_{\odot}}$ \citep{Shibata19}. 

EM observations of the kilonova can also be used to inform our inferences on $\tilde \Lambda$ values derived from GW data. Expanding on work by \cite{Radice18JointConstraint}, \cite{RadiceDai19} infer $\tilde \Lambda\gtrsim 300$  based on a lower limit of $0.04\,\rm{M_\odot}$ on the merger remnant disk mass that is necessary to support accretion disk winds powerful enough to deposit the observed kilonova ejecta mass.  
Similarly, \cite{Coughlin18,Coughlin19multimessbayes} derive $\tilde \Lambda> 300$ and  $m_1/m_2\in (1,1.27)$ at 90\% confidence level.
The inclusion of EM inference in the analysis of \gw{} thus lowers the support around the BH-BH region of the parameter space ($\tilde \Lambda =0$). (Astrophysical support for BH-BH systems with the very small masses inferred for \gw{} is scarce based on our current understanding of stellar evolution). However, while a BNS merger is favored,  a joint EM+GW analysis leaves open the possibility of a BH-NS system (e.g., \citealt{Coughlin19BHNS, Hinderer19}) unless at least some of the blue kilonova material originated from shock-heated ejecta at the interface of the two colliding NSs or a neutrino-driven wind from a remnant HMNS. 
Finally, these hybrid EM+GW approaches also enable estimates of NS radii that are more precise than those based on pure GW  analyses (e.g., \citealt{Bauswein17} ).  For example, \cite{RadiceDai19} derive $R(1.4\,\rm{M_{\odot}})=12.2^{+1.0}_{-0.8}\pm0.2\,\rm{km}$ (90\% confidence, statistical and systematic uncertainties). The addition of nuclear physics constraints allowed \citet{capano20} to obtain $R(1.4\,\rm{M_{\odot}})=11.0^{+0.9}_{-0.6}\,\rm{km}$ (90\% confidence).

To summarize, these multimessenger 
endeavors are ultimately enabled by the fact that physically, the merger process, the post-merger remnant evolution, the mass-ejection process, and the ejecta mass properties (as well as the properties of the emerging ultra-relativistic jet) fundamentally depend on the NS EoS. At the time of writing, their major limitation  is related to the level of advancement of current models of the EM signal, which in most cases rely on a set of numerical relativity simulations that do not cover the entire parameter space (e.g., \citealt{Kiuchi19}). A larger sample of BNS mergers with EM+GW detections \emph{and} an improved quantitative understanding of the EM emission is necessary to realize the full scientific potential of multimessenger parameter estimation in future work. This is the current leitmotiv of  joint EM+GW studies of compact-object mergers.

\subsection{Precision cosmology with GWs and their EM counterparts} \label{SubSec:Cosmology}

\begin{table}[h]
\tabcolsep7.5pt
\caption{$H_0$ values derived from multimessenger analysis of \gw{}.  The first two rows list $H_0$ estimates based on GW data, the sky position of the EM counterpart and the Host-Galaxy (HG) redshift. The following three rows combine the inferences on the observing angle derived from the afterglow (\S\ref{Sec:Afterglow}). The CMB and local-distance ladder estimates of $H_0$ are also listed for ease of comparison.  }
\label{Tab:H0}
\begin{center}
\begin{tabular}{@{}l|c|c|c@{}}
\hline
Method &$H_{0}$ & Model & Reference \\
 & {(}$\rm{km\,s^{-1}Mpc^{-1}}$)$^{\rm b}$ & &  \\
\hline
GW+EM position+HG$^{\rm a}$ & $70^{+13}_{-7}$ & High-spin case & \cite{abbott19properties} \\
GW+EM position+ HG  & $70^{+19}_{-8}$ & Low-spin case & \cite{abbott19properties}\\
GW+EM position+ HG + $\theta_{obs}$ & $70.3^{+5.3}_{-5.0}$ & Hydrodynamical jet & \cite{Hotokezaka19}$^{\rm c}$\\
GW+EM position+ HG + $\theta_{obs}$ & $68.1^{+4.5}_{-4.3}$ & Power-law jet & \cite{Hotokezaka19}\\
GW+EM position+ HG + $\theta_{obs}$ & $68.3^{+4.4}_{-4.3}$ & Gaussian jet & \cite{Hotokezaka19}\\
GW+EM position+ HG + $\theta_{obs}$ & $69.5^{+4.0}_{-4.0}$ & MHD jet & \cite{Wang20}\\
\hline
Planck (CMB) & $67.74\pm0.46$$^{\rm d}$  & TT,TE,EE+lowP+lensing+ext & \cite{PlanckH0} \\
SH0ES (Ia SNe) & $73.24 \pm 1.74$ & -- & \cite{Riess16H0}\\
SH0ES+LMC Cepheids & $74.03 \pm 1.42$ & -- & \cite{riess19}\\
\hline
\end{tabular}
\end{center}
\begin{tabnote}
$^{\rm a}$ All the methods listed here assume a peculiar velocity $v_p= 310 \pm 170\,\rm{km\,s^{-1}}$ for NGC\,4993. Note however that \cite{hjorth17,Guidorzi17} independently estimate a larger uncertainty $\sigma_{v_p}\sim230-260\,\rm{km\,s^{-1}}$ that would contribute additional uncertainty to $H_0$ (\citealt{Abbott17H0}, Extended Data Fig. 2); $^{\rm b}$ $1\sigma$ credible intervals listed; $^{\rm c}$ Favored model; $^{\rm d}$ Planck base $\Lambda$CDM value.
\end{tabnote}
\end{table}

GW sources offer a standard siren measurement of $H_0$  \citep{Schutz86} that is independent of a cosmic distance ladder, does not assume a cosmological model as a prior, and is thus well positioned to resolve the current tension between the {\it Planck} and the Cepheid-supernova measurements of $H_0$ \citep{PlanckH0,riess19}.
The additional constraints provided by the 
redshift of the galaxy hosting the GW event have been explored and quantified by \citet{holzhughes05,Nissanke10,Nissanke13} in the context of $H_0$.

Using the sky position of \gfo and combining the recession velocity derived from the redshift (and peculiar velocity) of the host galaxy NGC\,4993  with the distance to the source derived from GW data, \cite{Abbott17H0} infer  $H_0=70.0^{+12.0}_{-8.0}\,\rm{km\,s^{-1}Mpc^{-1}}$ (1$\sigma$). A re-analysis of the GW data led \cite{abbott19properties} to slightly revised values listed in \textbf{Table \ref{Tab:H0}}. It is estimated that $\sim60-200$ GW-detected BNS mergers with identified host galaxies will lead to a precision measurement of $H_0$ at the level of $2\%-1\%$, respectively, (\citealt{Chen18}, see also \citealt{Feeney19}). At the time of writing this is one of the most promising venues to clarifying the discrepancy between existing local (i.e., Type Ia SNe) and high-$z$ (CMB) $H_0$ measurements.

For these multimessenger measurements of $H_0$ the major source of uncertainty is the degeneracy between the luminosity  distance and the inclination angle of the binary, which is intrinsic to GW data of BNS mergers  \citep{Abbott17H0,Chen19}.
The jet afterglow provides an independent measure of the observing angle  (\S\ref{Sec:Afterglow}). Assuming that the jet
is aligned with the binary rotation axis, this information can be used to solve the GW parameter degeneracy \citep{Guidorzi17}. Using this additional constraint on $\theta_{obs}$, the precision of the $H_{0}$ multimessenger measurement 
improves by a factor $\sim$$2$ \citep{Hotokezaka19,Wang20}. The accuracy of the method  relies on the modeling of the EM afterglow data and on specific assumptions to interpret the data (e.g., the assumed jet structure, modeling of the hydrodynamical jet spreading, etc.). These EM systematics currently constitute the major limitation of this method and dominate over the GW systematics, which are primarily related to  the instrumental calibration error in the amplitude of the signal (e.g., \citealt{Chen19}).  While adding a new source of systematics and being limited to bright afterglows in the nearby universe ($d\lesssim 100$ Mpc) where the afterglow can be imaged, detected, and well sampled, this method brings the benefit of significantly reducing the number of BNS mergers necessary for $1$\% precision measurements of $H_0$  to $\mathcal{O}(10)$ \citep{Chen19,Hotokezaka19}.

\subsection{Tests of General Relativity (GR) and Fundamental Physics} \label{SubSec:GRtests}
The joint detection of gravitational and electromagnetic waves from the same celestial body enables powerful new ways to test general relativity and fundamental physics. The measured delay between GWs and $\gamma$-rays of  $\Delta t_{GW-\gamma}=$\delayGammaRaysGBM (\S\ref{SubSec:GammaRays170817}) constrains the difference between the speed of gravity $c_g$ and speed of light $c$  as  $-3\times 10^{-15} \le (c_g/c -1 ) \le 7 \times10^{-16}$ \citep{abbott17gammaray,Shoemaker18}. This multimessenger measurement places new bounds on the violation of Lorentz invariance and offers a new test of the equivalence principle by probing whether EM and gravitational radiation are equally affected by the background gravitational potential, which can be quantified with an estimate of the Shapiro delay \citep{abbott17gammaray,Wei17,Boran18}. The tight $|c_g/c-1|$ constraint severely limits the parameter space of modified theories of gravity that offer alternative gravity-based explanations to dark energy or are dark matter ``emulators'' \citep{Baker17,Creminelli17,Sakstein17,Ezquiaga17,Langlois18,Boran18,Dima18,Rham18}.  Additionally, the comparison between the EM and GW distance of \gw{} allows inferences to be drawn on the presence of additional large spacetime dimensions, which are found to be consistent with the GR prediction of $D=4$ \citep{Pardo18,Abbott19testsGR170817}. Finally, the NS coalescence GW signal enables tests on the deviation from the general-relativistic dynamics of the source and on the propagation and polarization of GWs \citep{Abbott19testsGR170817}. No significant deviation from GR expectations has been found.

\section{ENVIRONMENT}
\label{Sec:environment}
\subsection{Host galaxy properties and properties of the local environment}
\label{SubSec:HG}

\begin{figure}[h]
\includegraphics[scale=0.45]{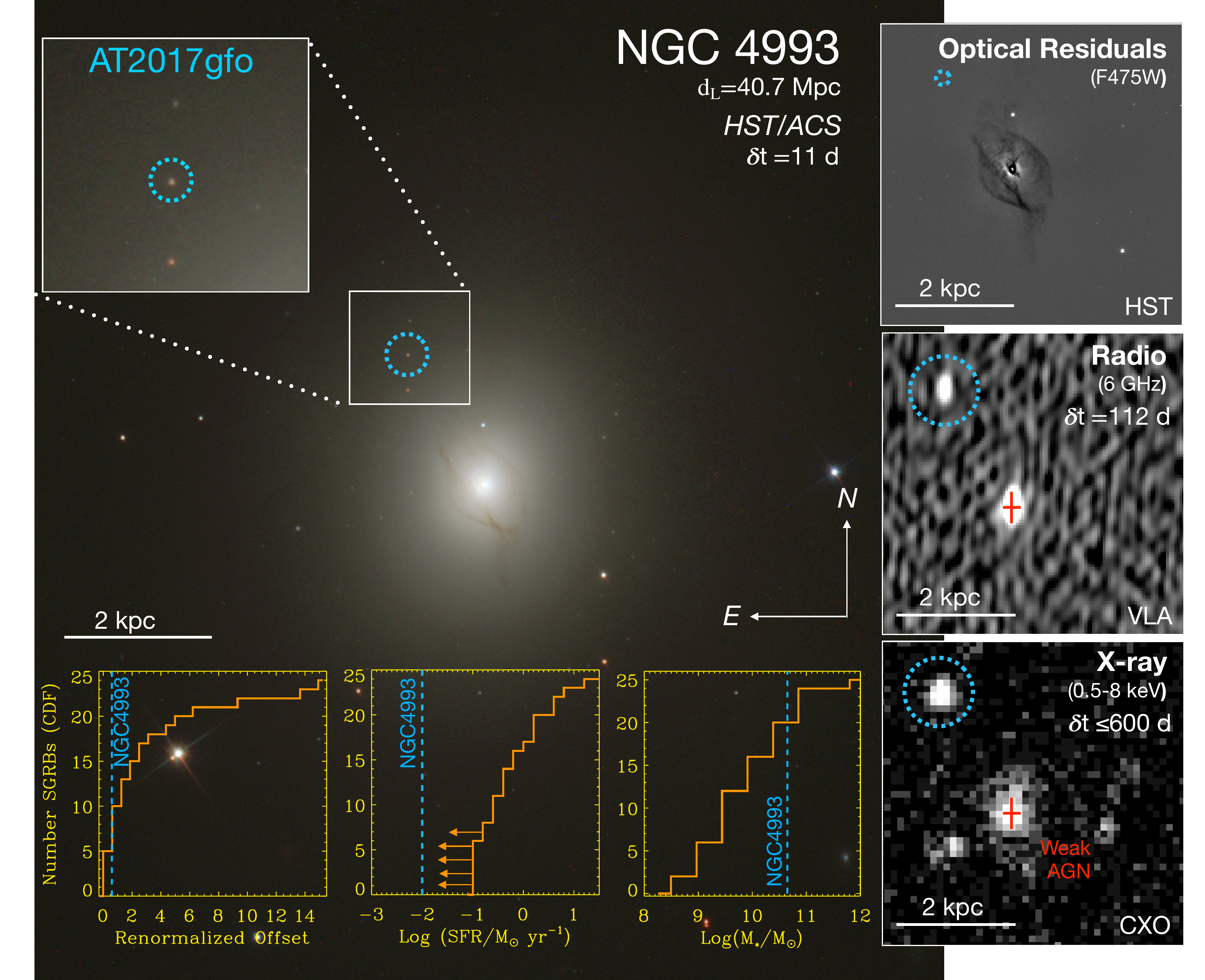}
\caption{\emph{Left, main panel}: Color image of NGC\,4993 created from filtered HST/ACS images (F850LP, F625W, F475W) acquired at $\delta t\approx11$ d by \cite{Blanchard17}. The location of the EM counterpart to \gw{} is identified by a dashed light-blue circle. \emph{Lower panels}: Inferred properties of NGC\,4993 (\S\ref{Sec:environment}) compared to the renormalized offset, star formation rate (SFR) and stellar mass distributions  of host galaxies of SGRBs from \cite{Leibler10,Berger14,Fong17,Nugent20}. \emph{Right panels,} from top to bottom: residual image in the HST
ACS/F475W filter after the subtraction of a S{\'e}rsic brightness profile with $n\approx 4$ showing the presence of dust lanes around the host nucleus (from \citealt{Blanchard17}). Middle and bottom panels: radio (VLA, 6\,GHz) and X-ray (CXO, 0.5--8 keV) image of the field acquired when the jet afterglow is clearly detectable. X-ray and radio emission from a weak AGN is also present and marked with a red cross (observations published by \citealt{Alexander18GW,Margutti18GW,Hajela19}).  }
\label{Fig:HG}
\end{figure}

\gw{} was localized at a projected offset of \offsetarcsec\, (\offsetkpc) from the center of the S0 galaxy NGC\,4993 (\textbf{Figure \ref{Fig:HG}}; e.g., \citealt{Blanchard17,Im17,Kasliwal17,Levan17,Palmese17,Pan17}). NGC\,4993 is well modeled by an $n\approx 4$ S{\'e}rsic profile and shows a strong bulge component as well as a complex morphology of 
dust lanes, concentric shells, and spiral features that indicate a relatively recent (\HGmergerepoch) galaxy merger. The global SED of NGC\,4993 has been fitted with multiple methods, assumptions and data sets (e.g., \citealt{Blanchard17,Pan17},) 
leading to a concordant picture of a massive galaxy (stellar mass $\log(M_{*}/M_{\odot})=$\HGstellarmass) with an old stellar population (half-mass assembly time  $\tau_{1/2}=$\HGstellarage) and limited ongoing star formation ($\log(\rm{SFR}/M_{\odot}yr^{-1})=$\HGSFR).
Due to higher SFR at early times, $90\%$ of the total stellar mass was formed by $\sim$  \HGlastSFRepoch\, ago. 
Additionally, the emission-line properties are consistent with the presence of a weak active galactic nucleus (AGN) revealed by nuclear X-ray and radio emission  in significant excess of the inferred star formation (\HGXraylum\,, \HGradiolum, \citealt{Blanchard17,Levan17}). The early-type morphology of NGC\,4993 is similar to $\sim 1/3$ of SGRB host galaxies and the presence of AGN activity is not unprecedented among SGRB host galaxies.  NGC\,4993 is superlative among SGRB host galaxies because of the very old stellar population age and exceedingly low SFR \citep{Fong17}.

Next, we constrain the properties of the BNS merger's local environment with optical, radio, and X-ray observations of NCG\,4993. 
\gw{} is located within the half-light radius of its host galaxy ($R\sim$\offsetrenormalized$r_e$, where $r_e$ is the S{\'e}rsic effective radius; \citealt{Blanchard17,Levan17,Pan17}), at a relatively bright location and relatively close to the host-galaxy center. As a comparison, $\approx 75\%- 80\%$ of SGRBs are localized in fainter regions of their host galaxies and $\approx$25\% of SGRBs are more proximal to their host centers (\textbf{Figure \ref{Fig:HG}}; \citealt{Fong17}).  Constraints on the local environment density of \gw can be placed by radio observations that probe the HI surface density and by X-ray observations of bremsstrahlung emission from hot plasma in the host galaxy, which probe the presence of ionized H. \cite{Hallinan17} infer a local number density of neutral H, $n_{\rm{HI}}<0.04\,\rm{cm^{-3}}$, while \cite{Hajela19} infer $n_{H^{+}}<0.01\,\rm{cm^{-3}}$ (3$\sigma$ c.l.) from diffuse X-ray emission at the location of \gw (consistent with the less constraining limit on $n_{H^{+}}$ reported by \citealt{Makhathini2020}). Deep HST observations acquired at $\delta t\approx\,$584 d (after the afterglow had completely faded away) rule out pre-existing emission from point sources with absolute $M_{F606W}< -4.8$ mag at the transient location \citep{Fong19}. These findings offer independent support to the low-density environment suggested by the afterglow studies  (\S\ref{SubSec:JetAfterglow}).

The global properties of NGC\,4993 and the properties of the transient location are thus consistent with those of the population of SGRBs, and in line with expectations for BNS mergers.

\subsection{Implications on the progenitor formation }
\label{SubSec:progenitor}
We explore the inferences on the progenitor formation of \gw that can be drawn from the 
properties of its global and local environment (\S\ref{SubSec:HG}).  The final BNS merger location depends on a combination of factors (e.g., \citealt{Abbott17progenitor}) including (i) the location of initial formation and the binary formation channel (e.g., dynamical vs. binary stellar evolution); (ii) initial binary properties, which determine the binary evolutionary path and the NS masses; (iii) systemic velocity of the binary after the second supernova (SN) explosion, which imparts a SN kick to the newly formed NS and a mass-loss kick on the companion NS; (iv) host-galaxy gravitational potential where the binary moves.

There is no observational evidence that supports a dynamical formation scenario of the \gw{} progenitor in a globular cluster (GC, \citealt{Blanchard17,Levan17,Pan17,Lamb19}). Deep {\it HST} observations obtained at $\delta t\approx\,$584 d place a luminosity limit $\lesssim 6.7\times 10^{3}L_{\odot}$ ($M_{F606W}\gtrsim -4.8$ mag) on any GC at the location of GW\,170817 \citep{Fong19}, making in situ formation of the BNS in a GC very unlikely (only $\approx 0.004\%$ of the total mass in GCs in NGC\,4993 is below this limit). Progenitor formation within a GC and a later ejection before merger (e.g., \citealt{Andrews19}) cannot be ruled out, as that would require the capability to correlate GW\,170817 with its parent GC after a long merger timescale. 

The very old stellar population of NGC\,4993 indicates long BNS merger timescales ($>1\,\rm{Gyr}$; \citealt{Blanchard17,Levan17,Pan17,Abbott17progenitor}). The combination of a long inferred merger timescale and relatively small projected offset suggests that the binary experienced a modest SN kick. Binary population synthesis modeling that used as inputs the NS mass posteriors of \gw{} derived from GW data, the measured projected offset, and the inference of the old stellar population of NGC\,4993 derived a second SN kick of $\approx 300^{+250}_{-200}\,\rm{km\,s^{-1}}$, assuming that the \gw{} progenitor evolved as an isolated binary. However, observations of this single system cannot rule out the possibility of a kick fortuitously directed along the line of sight (which would minimize the projected offset), or  a binary on an extended orbit that merged close to the host-galaxy center (\citealt{Abbott17progenitor,Levan17}). Finally, the density in the merger's surroundings depends on whether the binary system has hosted a pulsar. A comparison to BNS systems in the Galaxy that are expected to merge within a Hubble time shows that in most cases the low-density cavity carved out by the pulsar winds extends to large radii that are not consistent with the early onset of the  afterglow of \gw \citep{Ramirez-Ruiz19}

Until the present day, the host-galaxy demographics and the predictions from binary stellar evolution have been used to support the case of SGRBs as products of NS mergers (e.g., \citealt{Berger14}; \S\ref{SubSec:thread3SGRBs}). In the next decade, with a statistical sample of GW-detected BNS mergers well localized by their EM counterparts, the flow of inference can be reversed, and BNS mergers can be used to inform binary stellar evolution models \citep{Levan17}.

\section{OPEN QUESTIONS AND FUTURE PROSPECTS}

\subsection{Future Observations of \gw: the kilonova afterglow}   \label{SubSec:KNafterglow}

X-ray observations of \gw at $\delta t$$>$$600$ d point at a possible flattening of the light curve (\textbf{Figure \ref{Fig:afterglow}}).  A number of factors could lead to this interesting (yet-to-be statistically significant) X-ray flattening. This effect could originate from the jet dynamics (i.e., the jet hydrodynamical spreading and/or deceleration into the non-relativistic (NR) phase,  
an overdensity encountered by the blastwave, or possibly the emergence of an additional emission component (e.g., \citealt{Granot18offaxislightcurve,Ryan20}). For the jet-environment parameters of GW\,170817, the full transition to the NR regime and the appearance of the counter jet is expected at 
$t_{\rm{NR}}$$\gtrsim$$ 3000$ days, 
and $\nu_c\gg$ X-rays at the present epoch. While the presence of an overdensity or variation in the shock microphysical parameters (e.g., $p$) cannot be excluded, perhaps the most interesting interpretation would be the emergence of non-thermal synchrotron emission from the deceleration of the kilonova ejecta into the environment: the kilonova afterglow \citep{Nakar11,Granot18offaxislightcurve,Kathirgamaraju2019KN,Margalit20}.

The potential emergence of the kilonova afterglow of \gw in the X-rays has been discussed by \citet{Hajela19,Troja20}. Similar to supernovae, the bulk of the kinetic energy in kilonovae is carried by ``slowly'' moving ejecta that power the 
UV-optical-NIR thermal emission (\S\ref{Sec:Thermal}). The significantly lighter kilonova fastest ejecta rush ahead and shock the medium, producing synchrotron emission that is expected to peak on timescales of $\gtrsim$yrs. 
The kilonova afterglow maps the emission from this fast tail of ejecta with $v\gtrsim 0.3$c, and constrains
the ejecta kinetic energy structure 
in the velocity space $E_k^{\rm{KN}}(\Gamma \beta)$.  $E_k^{\rm{KN}}(\Gamma \beta)$ carries direct information on the merger dynamics, on the presence of the very fast tail of ejecta that is invoked by cocoon shock breakout  models to produce \grb (\S\ref{SubSec:GammaRays170817}), 
and, potentially, on the nature of the compact object
remnant (\S\ref{SubSec:RemnantNature}; e.g., \citealt{Hotokezaka18,Radice18viscous,Radice18BNS,Fernandez19}).

While limited statistical evidence for the emergence of a kilonova afterglow currently exists in the X-rays, the kilonova afterglow is expected to be more prominent in the radio domain because of the location of the synchrotron frequency. With the improved sensitivity of the next generation of X-ray and radio observatories (e.g., ngVLA, SKA1-MID, Lynx),
it will be in principle possible to detect the kilonova afterglow of \gw for decades  \citep{Alexander17}. 

\subsection{Nature of the compact-object remnant} \label{SubSec:RemnantNature}
GWs from the post-merger phase are the only direct probe of the nature of the merger remnant object, which might be a BH or NS, with massive NSs above the stability thresholds imposed by the nuclear matter EoS eventually collapsing to a BH on different timescales (e.g., \citealt{fryer15}, see \citealt{Bernuzzi20} for a recent review).  
A search for post-merger GWs from \gw on short ($\lesssim 1$ s) and intermediate-duration ($\lesssim 500$ s) timescales in the kHz regime, where hypermassive and supramassive NS remnants are expected to radiate, led to no significant detection in Advanced LIGO, Advanced Virgo and GEO600 data  \citep{Abbott17postmerger,abbott19properties}.  The derived limits do not constrain the direct BH collapse scenario, as the remnant BH ringdown GW signal is expected to be significantly below the threshold for current detectors \citep{Abbott17postmerger}. However, the post-merger GW limits interestingly lie within a factor $\sim 10$ from the theoretically predicted range of GW strain amplitudes from hypermassive and supramassive NS remnants, which might be thus meaningfully probed in the future by the full LIGO-Virgo network at design sensitivity \citep{abbott19properties}.
\begin{marginnote}[]
\entry{Stable NS}{NS with mass below the Tolman-Oppenheimer-Volkoff (TOV) mass $M_{TOV}$.}
\end{marginnote}
\begin{marginnote}[]
\entry{$M_{rot}^{NS}$} {mass limit for a uniformly rotating NS.}
\end{marginnote}
\begin{marginnote}[]
\entry{Supramassive NS}{NS with $M_{TOV}<M<M_{rot}^{NS}$ that will spin down through emission of GWs and light, and collapse to BH over $\sim10-5\times 10^{4}$ s.}
\end{marginnote}
\begin{marginnote}[]
\entry{Hypermassive NS}{NS with $M>M_{TOV}$ and $M>M_{rot}^{NS}$, that is  temporarily stabilized against gravitational collapse for $\sim$1 s by differential rotation and thermal gradients. 
}
\end{marginnote}
The nature of the merger remnant leaves several potential imprints on the EM counterpart that fall under four major categories (e.g., \citealt{Bauswein13,Metzger14,MargalitMetzger17,Murguia-Berthier17,Shibata17,Margalit19}): (i) kilonova colors; (ii) amount of ejecta mass; (iii) ejecta kinetic energy; (iv) presence of a successful relativistic jet. All these signatures fundamentally depend on the lifetime of the NS remnant, with bluer, more massive, and more energetic kilonovae without a successful jet pointing at longer-lived NSs.   While EM observations of \gw offer no conclusive evidence and a long-lived or stable NS cannot be entirely ruled out on EM and GW grounds (e.g., \citealt{abbott17gammaray,abbott20model,Piro19,Troja20}), the presence of a blue kilonova component associated with a large mass of lanthanide-free ejecta and $E_k\sim10^{51}\,\rm{erg}$ (\S\ref{Sec:Thermal}) together with a successful relativistic jet (\S\ref{Sec:Afterglow}) strongly disfavors a prompt collapse to a BH, and argues in favor of a hypermassive NS that collapsed to a BH within a second or so after merger (\citealt{Granot17,MargalitMetzger17,Shibata17,Metzger18magnetar,Rezzolla18,Gill19BHcollapse,Ciolfi20MNRAS,Murguia-Berthier20}). Finally, there is no 
significant EM observational evidence for long-lived ($t>T_{90}$) central engine activity in the form of accretion onto a BH or spindown of a rapidly rotating NS (\citealt{abbott17gammaray,Pooley18,Hajela19}, \S\ref{SubSec:othersources}). 

EM and GW observations of \gw strongly motivate improvements to the high-frequency sensitivity of GW interferometers to directly probe the outcome of the merger and accurately map EM signatures to the properties of their compact remnants.


\subsection{Early optical observations}
\gfo\ was not detected in the optical until 10.9~hr after the merger \citep{coulter17}, mostly due to its unfavorable sky location. The emission at blue wavelengths was unexpectedly luminous for an \rp-powered kilonova and several alternative models have been proposed (\S \ref{SubSec:BlueEmission}). Although the theoretical models will be refined, distinguishing between them will ultimately require high-cadence multicolor observations in the first day after a future BNS merger \citep[e.g.,][]{Arcavi18}. 
 Even in a standard blue kilonova model, the shape of the light curve in the first day is a sensitive probe of the structure of the outer ejecta \citep{Kasen17,banerjee20}. 
At sufficiently early times, additional components such as free neutron decay have been proposed to contribute to the observed emission \citep{kulkarni05,metzger15neutron, Gottlieb20}. 
Polarimetry will also be a powerful tool to study the ejecta geometry, particularly in the first day or two after the merger (\S \ref{sec:pol}; \citealt{bulla19}). 

\subsection{Is there a role for other production sites for the \rp? }
The observations discussed in \S \ref{subsec:rprocess} demonstrate that \gw\ ejected about the right amount of \rp-enhanced material to explain Galactic nucleosynthesis if it is typical of BNS mergers \citep[e.g.,][]{rosswog18}.  There is room for substantial future refinements in the estimates of the BNS merger rate and in the \rp production of individual events.
 Nevertheless, it is still possible that extreme or unusual supernovae  \citep[e.g.,][]{siegel19collapsar} contribute to the observed abundances, particularly for the first \rp peak, where observations of metal-poor stars indicate some scatter relative to the solar abundance pattern \citep{sneden08,cowan19}. 
 Also, there appears to be emerging evidence that there is a substantial dispersion in the observed properties of the kilonovae associated with SGRBs (\S \ref{SubSec:KNeSGRBs}).
\citet{cote19} and \citet{ji19} have comprehensively reviewed the connection between nucleosynthesis in NS mergers and the abundance patterns in metal-poor stars. Their reviews of the abundance ratios find a shortage of evidence that \gw ejected sufficient amounts of the heaviest \rp material relative to the lighter \rp. However, material with the highest \xlan might only manifest at the latest times and in the IR (\S \ref{subsubsec:lateobs}). {\it JWST} and the planned 20--30~m ground-based telescopes will be necessary to obtain the required observations in the IR at late times for future NS mergers.
 
\subsection{Neutrinos} \label{SubSubSec:neutrinos}
The three most sensitive high-energy neutrino observatories (i.e., ANTARES, IceCube, and Pierre Auger) searched for GeV--EeV neutrinos associated with \gw and reported no evidence for directionally coincident neutrinos within $\pm500\,\rm{s}$ around the merger time and up to 14~d post-merger \citep{Albert17neutrinosdata}. A search for MeV neutrinos with IceCube similarly led to a non-detection, consistent with the observed properties of GRB\,170817A, and the expectations from SGRB jets and their extended $\gamma$-ray emission, on-axis or off-axis. Prompt neutrinos (i.e., neutrinos associated with the prompt GRB emission) would reveal the hadronic content of the jet, provide insight into particle acceleration and the dissipation mechanism in relativistic outflows, with the added benefit of an improved localization of the GW event \citep{Albert17neutrinosdata}. The detection of a long-lived source of high-energy neutrinos (timescales of $\sim$ days) would additionally point to an equally long-lived or indefinitely stable millisecond magnetar merger remnant (e.g., \citealt{Fang17}). 

\subsection{Searches for kilonovae untriggered by GW or GRB detections} \label{SubSec:untriggeredSearches}
Blind searches for kilonovae as optical transients untriggered by SGRBs or GWs  
have been carried out before and after the discovery of AT\,2017gfo using data streams from a variety of telescope surveys
\citep{Doctor17,Kasliwal17,Smartt17,Yang17,Andreoni20,McBrien20}.
No kilonova has been confidently identified so far, with only one potential candidate reported by \cite{McBrien20}. However, the null results from these searches can be used to independently constrain the rate of kilonovae in the local Universe. The tightest limit on the rate of kilonovae with luminosity similar to AT\,2017gfo is $\mathcal{R}<800\,\rm{Gpc^{-3}yr^{-1}}$ (\citealt{Kasliwal17}; also see Fig. 9 of \citealt{Andreoni20}). A direct comparison to the rates of SGRBs and BNS mergers derived from GWs of \textbf{Figure \ref{Fig:Rates}} might not be justified as we have clear evidence of kilonovae that are fainter than AT\,2017gfo (\S \ref{SubSec:KNeSGRBs}). The major challenges faced by all of these attempts stem from  the combination of the intrinsically low rate and faintness of the expected kilonova events, the limited volume of the Universe probed, and the comparatively  large rate of contaminants. Increasing the depth of these searches is mandatory to reveal the population of jet-less BNS mergers (if there), which might go undetected in the absence of a GW trigger.

\section{CONCLUSIONS} \label{Sec:Conclusions}
The first joint detection of GWs and light from the same celestial body offered an unprecedented opportunity for advancement in astrophysics, fundamental physics, gravitation, and cosmology. Like few other astronomical events, \gw and its EM counterparts tied together many threads of previous inquiry.
The thousands of scientists that participated in the enormous collective observational endeavor demonstrated a litany of firsts:

\begin{itemize}
\item First GW detection of a BNS merger
\item First secure discovery of an EM counterpart to a GW source
\item First  observations of a structured relativistic jet  seen from the side 
\item First definitive detection of a kilonova
\item First optical-NIR spectroscopy of a kilonova
\item First BNS merger to be localized in the local Universe
\end{itemize}

The interpretation of the observations relied on decades of theoretical work, with many of the key advances occurring only in the last few years before \gw. That the observations were so readily interpretable within this framework represents a triumph of theoretical astrophysics.

The implications of these studies of \gw bridge multiple disciplines:
\begin{itemize}
\item The properties of  \gw,  \grb,  and the non-thermal afterglow provide the first direct connection between SGRBs and BNS mergers. \gw demonstrated that BNS mergers can successfully launch highly collimated ultra-relativistic jets with core properties similar to those of SGRBs and substantial angular and radial structure that was simply impossible to probe before this event. \gw provided the first opportunity to appreciate that such a structure exists and has observable consequences.
\item Photometric and spectroscopic observations of \gfo in the optical and IR established that BNS mergers are cosmic sites of $r$-process nucleosynthesis, thus identifying the origin of (at least some of) the heaviest chemical elements of the periodic table.
 \item The small interval of time  
 between the GW coalescence and the detection of $\gamma$-rays enabled new tests of general relativity and fundamental physics, which, among other results, led to severe constraints on a sizeable fraction of the parameter space of modified theories of gravity. 
 \item GW measurements enabled direct constraints on the EoS of dense nuclear matter, which were strengthened by the combination with information from the EM counterpart. This multimessenger approach will be a major emphasis of future BNS investigations.
 \item Standard-siren based measurements of $H_0$ from EM-localized GW sources like \gw provide a novel method of inference in cosmology.  This method, put into practice with \gw for the first time, allows a measurement of  $H_0$ in the local Universe that does not rely on a cosmic distance ladder, and it is thus well positioned as one of the most promising venues to solve the current  $H_0$ tension.
 \end{itemize}

Against this backdrop of progress, \gw and its EM counterparts left key questions unanswered: 
(1) \emph{GWs:} What is the equation of state of dense nuclear matter? What is the ultimate merger remnant?
(2)\emph{Kilonova:} What is the origin of the blue emission component? Are BNS mergers the only significant site for the $r$-process? 
(3)\emph{Jets:} Do all BNS mergers successfully launch ultra-relativistic jets into their environments?  Is there a population of SGRBs powered by BH-NS mergers? Does jet launching require the collapse to a BH?  

This comprehensive dataset on \gw is unlikely to be equaled for any BNS merger in the near future. A critical question for future work is assessing how typical this event was of the whole population. There are hints from observations of SGRBs (\S \ref{SubSec:KNeSGRBs}) that the associated kilonovae exhibit substantial dispersion.
The future of the field also relies on improved multi-dimensional radiative transfer modeling, necessary to more quantitatively constrain nucleosynthesis and ejecta parameters in kilonovae, and improved MHD simulations of jets with realistic structures. 

Looking ahead, the increased sensitivity of future GW observatories will allow us to probe more distant populations of GW sources. However, GW interferometers are sensitive to the gravitational strain $h$, which scales as $1/d_L$ (where $d_L$ is the distance to a source), while optical telescopes are sensitive to the energy flux, which scales as $1/d_L^2$. The ultimate design sensitivity of Advanced LIGO will be sensitive to BNS systems out to $\sim$200 Mpc (and BH-NS star mergers at several hundred Mpc) and the proposed A$+$ upgrades by the 2025 time frame will aim to increase the sensitivity by an additional factor of $\sim$2. Therefore, planned technology developments in GW interferometers will rapidly start producing detections at distances where the expected optical KN counterparts are exceedingly dim, stretching the capabilities of even {\it JWST} and the next-generation Extremely Large Telescopes, and definitely beyond the reach of current X-ray, radio, and $\gamma$-ray facilities. For multimessenger astrophysics with GWs to flourish and keep its scientific promise, it is mandatory to match the upcoming generation of GW detectors with a new generation of more sensitive EM observatories across the spectrum.

\section*{DISCLOSURE STATEMENT}
The authors are not aware of any affiliations, memberships, funding, or financial holdings that
might be perceived as affecting the objectivity of this review. 

\section*{ACKNOWLEDGMENTS} 
We are grateful to the numerous observatory staffs around the globe who enabled the unprecedented observational campaign on this historic object.
The literature on this subject extends to thousands of publications, and we apologize in advance to those colleagues whose work we could not adequately discuss here.
We thank our long-time collaborators, including Edo Berger and Wen-fai Fong, for fruitful observational campaigns over the years and acknowledge the key role that discussion and models produced by Brian Metzger and Dan Kasen had in enabling our interpretation of the data.
It is our pleasure to acknowledge K. D. Alexander, D. A. Brown, P. Blanchard, L. DeMarchi, M. R. Drout, D. Finstad, G. Ghirlanda, O. Gottlieb, J. Granot, K. Hotokezaka, A. Kathirgamaraju, D. Lazzati, B. Margalit, S. Rosswog, and V.~A. Villar for their comments and for promptly replying to our requests and sharing data and results from simulations 
that have been used in figures of this review.
This research was supported in part by the National Science Foundation under Grant No. NSF PHY-1748958, No. AST-1909796 and AST-1944985. Portions of this work were initiated at the program on ``The New Era of Gravitational-Wave Physics and Astrophysics" (GRAVAST19) at the Kavli Institute for Theoretical Physics, UC Santa Barbara.   R.M. acknowledges partial support by NASA through Chandra Award Number G09-20058A and through Space Telescope Science Institute program \#15606. 
RC acknowledges partial support from Chandra grants DD8-19101B and GO9-20058B.
 
\bibliographystyle{ar-style2-fix.bst}
\bibliography{MC}

\end{document}